\documentclass[%
 reprint,
 amsmath,amssymb,
 aps,
]{revtex4-2}

\usepackage{hyperref}
\usepackage[utf8]{inputenc}
\usepackage{newunicodechar}
\newunicodechar{ }{\,}

\usepackage{graphicx}
\usepackage{dcolumn}
\usepackage{amsmath}
\usepackage{supertabular}
\usepackage{multirow}
\usepackage{indentfirst} 

\usepackage{lineno}

\usepackage{color}
\usepackage{cancel}
\usepackage{ulem}
\usepackage{array}
\usepackage{tabularx}

\usepackage{amssymb}
\usepackage{xcolor}

\usepackage{setspace}
\makeatletter
\def\section{\@startsection{section}{1}{\z@}%
  {-3.5ex \@plus -1ex \@minus -.2ex}%
  {2.3ex \@plus.2ex}%
  {\normalfont\large\bfseries\centering}}
\makeatother

\begin{document}

\title{Open charm production and $\Lambda_{c}^{+}/D^{0}$ ratio in pp and Au+Au collisions at the RHIC}

\bigskip

\author{Bijun Fan}
\affiliation{Key Laboratory of Quark and Lepton Physics (MOE) and Institute of Particle Physics, 
Central China Normal University, Wuhan 430079, China}

\author{Chao Zhang}
\email[Corresponding author, ]{chaoz@whut.edu.cn}
\affiliation{School of Physics and Mechanics, Wuhan University of Technology, Wuhan, 430070, China}

\author{Liang Zheng}
\email[Corresponding author, ]{zhengliang@cug.edu.cn}
\affiliation{School of Mathematics and Physics, China University of Geosciences (Wuhan), Wuhan 430074, China}
\affiliation{Key Laboratory of Quark and Lepton Physics (MOE) and Institute
of Particle Physics, Central China Normal University, Wuhan 430079, China}	

\author{Shusu Shi}
\email[Corresponding author, ]{shiss@ccnu.edu.cn}
\affiliation{Key Laboratory of Quark and Lepton Physics (MOE) and Institute of Particle Physics, 
Central China Normal University, Wuhan 430079, China}

\date{\today}
\begin{abstract}

We study open charm hadrons production in pp and Au+Au collisions at $\sqrt{s_{\mathrm{NN}}} = 200$~GeV using an improved a multi-phase transport (AMPT) model.
Specifically, we show the transverse-momentum spectra and nuclear modification factors $R_{\mathrm{AA}}$ of $D^{0}$ mesons and $\Lambda_{c}^{+}$ baryons, as well as the $\Lambda_{c}^{+}/D^{0}$ ratio in pp and Au+Au collisions. 
The results obtained from the AMPT model simulations are compared with the STAR experimental data and found to be consistent. 
We further investigate the $\Lambda_{c}^{+}/D^{0}$ ratio by evaluating contributions from coalescence, fragmentation, and the combined coalescence+fragmentation mechanisms, and we find that fragmentation alone underestimates the pronounced enhancement in Au+Au relative to pp at low and intermediate $p_{\mathrm{T}}$, whereas the coalescence+fragmentation mechanism reproduces the observed trend significantly better. 
These results indicate that coalescence plays a key role in charm baryon productions and helps constrain the relative importance of different hadronization mechanisms in the ultra-relativistic nuclear collisions.

\end{abstract}

\maketitle

\section{Introduction}\label{sec.I}

Relativistic heavy-ion collisions at RHIC and the LHC create strongly interacting matter at extreme temperatures and energy densities, where QCD predicts a transition from hadronic matter to a deconfined quark-gluon plasma (QGP) \cite{Gyulassy:2004zy,PHENIX:2004vcz,STAR:2005gfr,Chen:2024aom,Luo:2020pef,Bzdak:2019pkr}. In this medium, quarks and gluons are no longer confined inside hadrons, but propagate as quasi-free degrees of freedom over a finite lifetime. 
Heavy quarks are predominantly produced in initial hard scatterings before the QGP is formed and can be hardly generated via thermal production in the medium \cite{Muller:1992xn,Lin:1994xma}. They experience the entire space-time evolution of the fireball and lose energy through interactions with the deconfined partons. 
As a consequence, the momentum distributions of open charm hadrons encode both the strength of the charm-medium interaction and the mechanism by which charm quarks hadronize.
Open charm hadrons such as $D^{0}$ mesons and $\Lambda_{c}^{+}$  baryons thus provide valuable probes of QGP properties \cite{vanHees:2005wb,Brambilla:2010cs,Andronic:2015wma}, including the mechanisms of heavy quark energy loss and the dynamics of charm hadronization in the expanding medium.

In recent years, extensive measurements of open-charm hadrons at RHIC have provided important information on how charm quarks interact with the medium \cite{STAR:2012nbd,STAR:2014wif,STAR:2018zdy,STAR:2019ank,STAR:2004ocv,STAR:2021tte,STAR:2017kkh}. 
In particular, the STAR Collaboration has reported transverse momentum spectra and nuclear modification factors $R_{\mathrm{AA}}$ of $D^{0}$ mesons in pp and Au+Au collisions at $\sqrt{s_{\mathrm{NN}}} = 200$~GeV over a broad range of centralities and $p_{\mathrm{T}}$ \cite{STAR:2012nbd,STAR:2014wif,STAR:2018zdy}. 
These data exhibit a characteristic suppression of $D^{0}$ yields at intermediate and high $p_{\mathrm{T}}$ in central Au+Au collisions relative to binary-scaled pp results, indicating substantial energy loss and transport of charm quarks in the QGP. 
However, while the $R_{\mathrm{AA}}$ of mesons effectively constrains the charm spatial diffusion coefficient and energy loss parameters, it is only weakly sensitive to the details of charm hadronization, thereby offering limited discrimination between whether charm quarks hadronize predominantly via fragmentation or through coalescence with light quarks from the medium. 
To fully understand charm dynamics, particularly at low and intermediate $p_{\mathrm{T}}$, one must also disentangle the effects of transport in the QGP from the subsequent hadronization process.

The production of charmed baryons offers an additional constraint to the charm quark evolution and its hadronization in the hot dense nuclear medium. STAR has measured $\Lambda_{c}^{+}$ production in Au+Au collisions at $\sqrt{s_{\mathrm{NN}}} = 200$~GeV, including its transverse momentum spectra and the $\Lambda_{c}^{+}/D^{0}$ ratio \cite{STAR:2019ank} in a limited kinematic range, while corresponding $\Lambda_{c}^{+}$ measurements in pp collisions at the same energy are not yet available. 
Because charm baryons are sensitive not only to in-medium energy loss but also to the details of the hadronization process, the combined systematics of $D^{0}$ and $\Lambda_{c}^{+}$ observables place nontrivial constraints on models of charm-quark dynamics in the QGP.
In elementary collisions such as $e^{+}e^{-}$ and pp \cite{Belle:2005mtx,LHCb:2013xam}, charm hadron production is usually described by vacuum-like fragmentation, which leads to a relatively small $\Lambda_{c}^{+}/D^{0}$ ratio. 
In contrast, in nucleus-nucleus collisions the presence of a hot and dense partonic medium opens the possibility for charm quarks to hadronize via coalescence with thermal light quarks. 
In such coalescence-type scenarios, the probability to form a charm baryon can be significantly enhanced at low and intermediate transverse momentum, leading to a sizable increase of $\Lambda_{c}^{+}/D^{0}$ in A+A collisions compared to pp \cite{STAR:2019ank,ALICE:2018hbc,ALICE:2021bib}. 
The observed enhancement of $\Lambda_{c}^{+}/D^{0}$ in Au+Au collisions at RHIC therefore suggests that charm hadronization in the QGP cannot be described by independent fragmentation alone, but instead reflects a nontrivial interplay between coalescence and fragmentation in the expanding medium \cite{STAR:2019ank}.

On the theoretical side, several perturbative and transport frameworks have been developed to describe heavy-flavor production in high energy collisions. In proton-proton and proton-nucleus systems, heavy quark cross sections are commonly computed in fixed-flavor-number schemes (FFNS) \cite{Mangano:1991jk} and their extensions, such as general-mass variable-flavor-number schemes (GM-VFNS) \cite{Kniehl:2004fy,Helenius:2018uul} and the fixed-order plus next-to-leading-logarithm (FONLL) formalism \cite{Cacciari:2005rk}, which provide a broadly successful description of open charm measurements, although the central FONLL predictions often undershoot the data. 
In nucleus-nucleus collisions, where a hot and dense QGP is formed, the in-medium evolution of heavy quarks is typically modeled using transport approaches based on Fokker-Planck \cite{Das:2010tj,He:2012df,Cao:2015hia,Lang:2012nqy} or relativistic Boltzmann equations \cite{Fochler:2010wn,Djordjevic:2013xoa,Xu:2015bbz,Song:2015ykw,Cao:2016gvr}, which encode multiple scatterings and energy loss in the evolving medium. Such frameworks have been successful in reproducing key heavy-flavor observables, including nuclear modification factors and flow coefficients. Nevertheless, most perturbative and transport models focus on specific stages of the collision and do not capture the full collision dynamics from fluctuating initial conditions to the subsequent partonic transport and hadronic phase on an event-by-event basis. 
This limitation motivates the use of more comprehensive modeling approaches that can consistently incorporate heavy quark production, medium interactions, and hadronization within a unified dynamical framework.

To investigate charm-quark dynamics and hadronization in a unified setting that simultaneously describes the bulk medium and heavy flavors, it is useful to employ a microscopic transport model that follows all stages of the collision. 
In this work, we use an improved version \cite{Zhang:2022fum,Zhang:2024zga} of the string-melting A Multi-Phase Transport (AMPT) model \cite{Lin:2004en,He:2017tla,Zheng:2019alz,Zhang:2019utb,Lin:2021mdn}, which describes the system in terms of fluctuating initial conditions from HIJING \cite{Wang:1991hta}, a partonic cascade, quark coalescence, and a hadronic afterburner. 
In this framework, charm-anticharm pairs are produced in initial hard scatterings and subsequently propagate through the deconfined medium via Zhang’s Parton Cascade (ZPC) \cite{Zhang:1997ej}. 
Finally, the freeze-out charm quarks then hadronize through competing coalescence and independent fragmentation channels. 
This unified setup enables a consistent, event-by-event study of how partonic interactions in the QGP and the hadronization process jointly shape open-charm observables. 
Specifically, we calculate transverse momentum spectra and nuclear modification factors $R_{\mathrm{AA}}$ of $D^{0}$ mesons and $\Lambda_{c}^{+}$ baryons in pp and Au+Au collisions at $\sqrt{s_{\mathrm{NN}}} = 200$~GeV, and investigate the $\Lambda_{c}^{+}/D^{0}$ ratio with an emphasis on disentangling the relative contributions from coalescence and fragmentation in the QGP.

The remainder of this paper is organized as follows. Section~\ref{sec.II} details the improved AMPT framework, describing the initial heavy flavor production, the implementation of the Cronin effect, and the specific criteria used for the hybrid coalescence-fragmentation mechanism. In Sect.~\ref{sec.III}, we present the model results for $D^{0}$ and $\Lambda_{c}^{+}$ spectra and $R_{\mathrm{AA}}$, comparing them with the available STAR experimental data. This section also provides a detailed analysis of the $\Lambda_{c}^{+}/D^{0}$ ratio, isolating the kinematic regions dominated by different hadronization channels. Finally, a summary of our findings is provided in Sect.~\ref{sec.IV}.

\section{The AMPT model}\label{sec.II}

In this work, we employ an improved AMPT string melting (AMPT-SM) model to study open charm production in pp and Au+Au collisions at $\sqrt{s_{\mathrm{NN}}} = 200$~GeV. 
Several refinements have been implemented to improve the treatment of heavy-flavor production and hadronization in AMPT. Specifically, charm-anticharm pairs are explicitly extracted from the initial hard scatterings provided by HIJING, rather than being subjected to the generic string-melting procedure. In addition, an initial-state transverse momentum broadening (Cronin effect) is applied to the produced $Q\bar{Q}$ pairs\cite{Cronin:1974zm,Accardi:2002ik,Vitev:2006bi}. Finally, a hybrid heavy-quark hadronization scheme is implemented, featuring a competitive interplay between coalescence with thermal light quarks and independent fragmentation.

In the standard AMPT initial condition, minijet partons and excited strings produced in nucleon-nucleon collisions are converted into on-shell quarks and antiquarks via the string-melting mechanism \cite{Andersson:1983ia,Andersson:1983jt}. Heavy quarks, however, are expected to be produced predominantly in early-stage hard scatterings and thus should be treated differently from soft partons originating from string fragmentation. Accordingly, in this work charm-anticharm pairs are directly extracted from the HIJING \cite{Wang:1991hta} initial condition. Each charm quark is assigned a formation time $t_{\mathrm{F}} = E / m_{\mathrm{T}}^{2}$ \cite{Lin:2004en}, where $E$ and $m_{\mathrm{T}}$ denote the charm quark energy and transverse mass, respectively, after which it begins to participate in the partonic transport evolution.

\begin{figure}[!htbp]
  \centering   
  \includegraphics[width=0.5\textwidth]{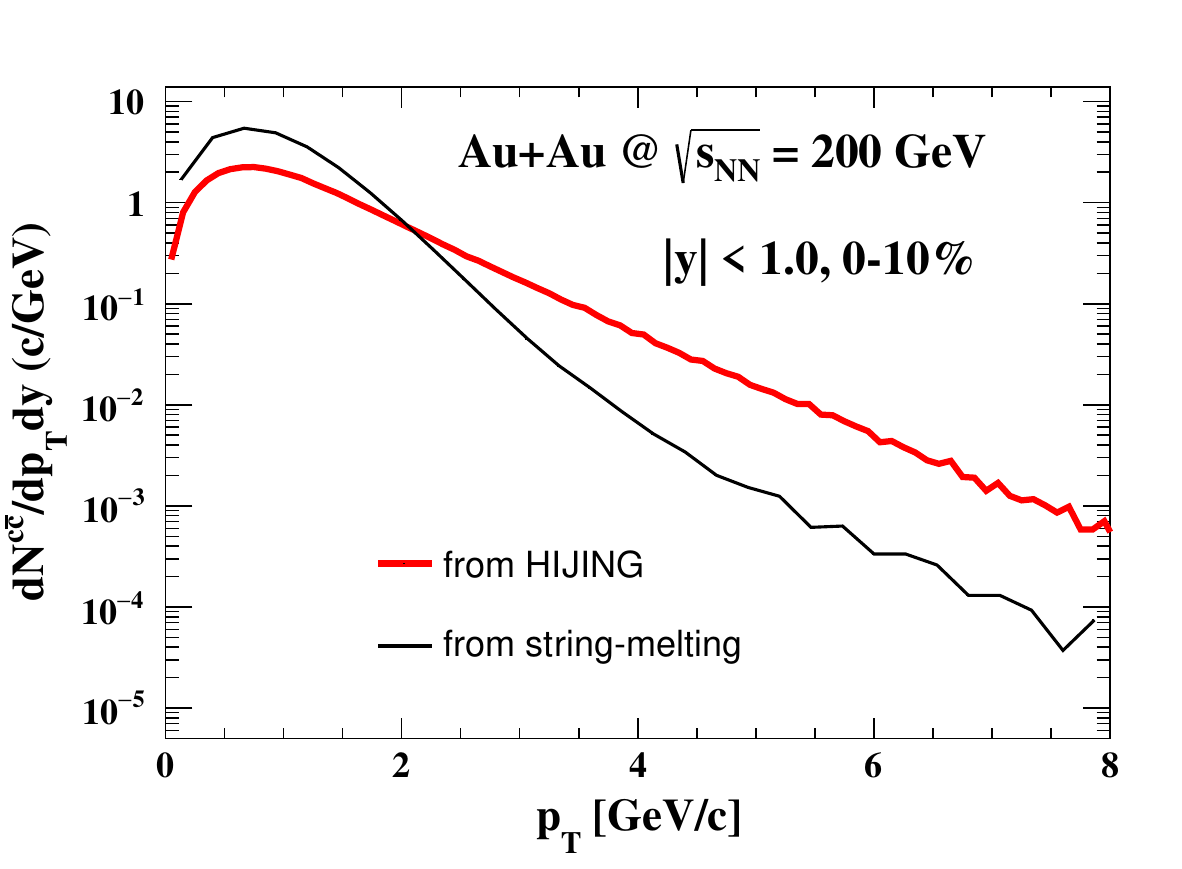}
   \caption{The $p_{\mathrm{T}}$ spectra of charm quarks at mid-rapidity in central Au+Au collisions extracted from the HIJING initial condition and from the string-melting process of the AMPT model at $\sqrt{s_{\mathrm{NN}}} = 200$~GeV.}
    \label{dNcc_dpt}
\end{figure}

Figure.~\ref{dNcc_dpt} shows the charm quark transverse momentum spectra in central Au+Au collisions at $\sqrt{s_{\mathrm{NN}}}=200$~GeV, The results from HIJING are compared with that from the string-melting mechanism.
The spectrum obtained from charm quarks taken directly from the HIJING initial condition is harder than that from the string-melting, consistent with the trend previously seen in small systems~\cite{Zhang:2024zga}. This difference reflects the distinct production mechanisms involved for the charm quarks. The charm quarks in HIJING originate predominantly from initial hard partonic scatterings, whereas the string-melting approach effectively redistributes the charm quark momentum through soft string fragmentation, leading to a softer spectrum. This comparison indicates that the initial treatment of charm production has a significant impact on the subsequent transport and hadronization of charm quarks.

\begin{figure*}[!htbp]
  \centering   
  \includegraphics[width=0.85\textwidth]{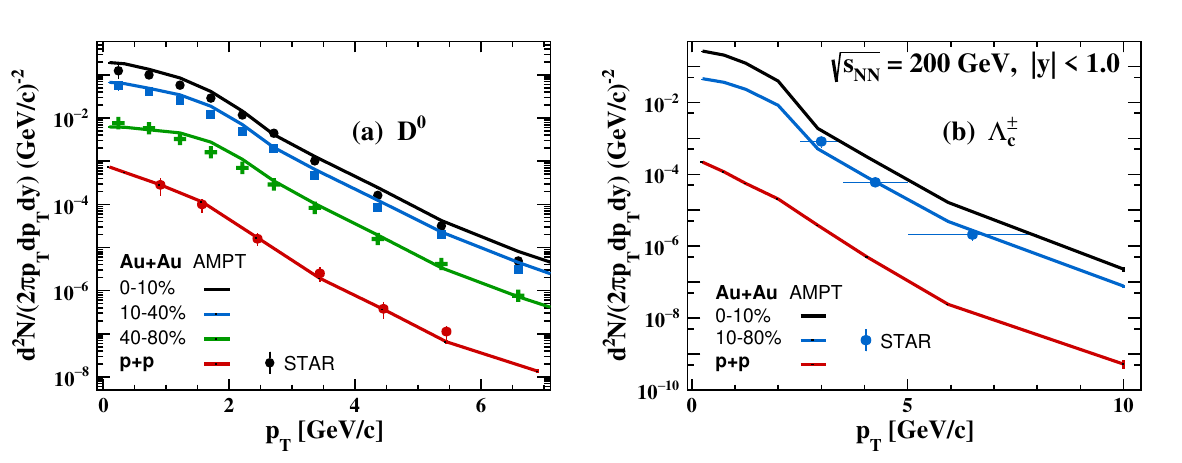}
   \caption{Transverse momentum spectra of (a) $D^{0}$ mesons and (b) $\Lambda_{c}^{\pm}$ baryons at mid-rapidity ($|y|<1.0$) in $pp$ and Au+Au collisions at $\sqrt{s_{\mathrm{NN}}}$ = 200~GeV. Model calculations from the updated string-melting AMPT setup are shown for $pp$ and for several Au+Au centrality classes, and are compared with the published STAR data where available. For $\Lambda_{c}^{\pm}$, experimental spectra are currently only available for 10--80\% Au+Au collisions. Model results are obtained with the same kinematic cuts and centrality selections as the measurements.}
   \label{D0_lc_ptspectra}
\end{figure*}

After partonic scatterings in ZPC cease, the system undergoes hadronization. In the standard version of AMPT, this stage is modeled exclusively through a spatial quark coalescence mechanism, applied to both light and heavy flavors.
However, for heavy quarks particularly at high transverse momentum, vacuum-like fragmentation is expected to play a significant role \cite{Cacciari:2002xb,MoosaviNejad:2013ssd}, as indicated by perturbative QCD studies of heavy quark fragmentation functions \cite{Mele:1990cw,Braaten:1994bz}. Therefore, we have implemented an independent fragmentation channel \cite{Sjostrand:1993yb} to complement the coalescence process.
In this hybrid scheme, charm quarks may hadronize either via coalescence with nearby light quarks or through independent fragmentation; in pp collisions, the coalescence contribution is naturally suppressed due to the much smaller partonic phase-space density. All freeze-out charm quarks are initially processed through the spatial coalescence algorithm to search for nearest-neighbor light quarks. However, to ensure physical kinematic matching, the charm quark and its pre-coalescence candidate partners are required to satisfy simultaneous constraints on their relative spatial separation and invariant mass,

\begin{equation}
d < p_{\mathrm{r}},
\label{eq:coal_pr}
\end{equation}
\begin{equation}
m_{\mathrm{inv}} < \sum m_{\mathrm{Q}} + p_{\mathrm{m}}\,\bigl(m_{\mathrm{H}}-\sum m_{\mathrm{Q}}\bigr),
\label{eq:coal_pm}
\end{equation}

where $d$ and $m_{\mathrm{inv}}$ are the relative distance and invariant mass of the quarks in their rest frame, $m_{\mathrm{Q}}$ denotes the constituent quark masses, and $m_{\mathrm{H}}$ is the mass of the pre-coalescenced hadron.
Charm quarks that do not meet the requirements above will hadronize via the independent fragmentation with the widely used Peterson fragmentation function \cite{Sjostrand:1993yb},

\begin{equation}
f(z) \propto \frac{1}{z\left(1-\frac{1}{z}-\frac{\epsilon_{\mathrm{Q}}}{1-z}\right)^{2}},
\end{equation}

where $z$ is the fraction of the heavy quark’s light cone momentum carried by the hadron, the parameter $\epsilon_{\mathrm{Q}}$ characterizes the hardness of the fragmentation and is typically set to 0.05 for charm quarks.

For pre-coalescence candidates that fail to satisfy Eqs.~\ref{eq:coal_pr} and ~\ref{eq:coal_pm}, the heavy quark $Q$ is separated from the rejected hadron candidate and reassigned to independent fragmentation. In this process, the heavy quark hadronizes according to the Peterson function by acquiring a light quark configuration from a vacuum-generated pair. The remaining recoil light quark component from the vacuum pair then combines with the original coalescence partner to undergo string fragmentation. This procedure ensures strict flavor conservation on an event-by-event basis.

In this work, the $p_{\mathrm{r}}=0.5~\mathrm{fm}$ and $p_{\mathrm{m}}=0.5$ are set for both pp and Au+Au collisions at $\sqrt{s_{\mathrm{NN}}}=200$~GeV.
These values are determined from a fit to the $D^{0}$ meson transverse momentum spectrum in pp collision at $\sqrt{s}=200$~GeV.
It has been noted that the AMPT-SM model produces hadrons too early in small systems like O+O collisions \cite{Zhao:2024feh}, resulting in the parton density at freeze-out exceeding the expected value for the QCD phase transition.
However, this behavior is expected to be weaker in larger systems such as Au+Au collisions due to stronger partonic interactions.
Considering that the coalescence effect should dominate charm hadronization at low transverse momentum, while fragmentation becomes increasingly important toward higher $p_{\mathrm{T}}$ \cite{Zhao:2023nrz,Hwa:2004ng,Song:2021mvc}, we introduce an additional selection criterion based on the hadron transverse-mass threshold to effectively separate low-density and high-density hadronization regimes associated with small and large collision systems

\begin{equation}
m_{\mathrm{T}} = \sqrt{m_{\mathrm{H}}^{2} + p_{\mathrm{T}}^{2}},
\end{equation}

where $m_{\mathrm{H}}$ denotes the mass of the produced hadron and $p_{\mathrm{T}}$ is its transverse momentum.
The pre-coalescenced heavy hadrons with $m_{\mathrm{T}}$ below the given threshold are assigned to the coalescence directly.
In practice, The parameter $m_{\mathrm{T}}=2.0$ for pp and $m_{\mathrm{T}}=3.3$ for Au+Au at $\sqrt{s_{\mathrm{NN}}}=200$~GeV. 
This selection effectively enhances the coalescence probability at low $p_{\mathrm{T}}$ in the nuclear environment and delivers a smooth transition from coalescence-dominated hadronization at low $p_{\mathrm{T}}$ to fragmentation-dominated production at high $p_{\mathrm{T}}$.

Following our previous work, the parameter $\delta = 1$ for the Cronin effect is constrained from the $D^{0}$ transverse momentum spectra in d+Au collisions at RHIC~\cite{STAR:2004ocv}.
The relative production of heavy flavor baryon to meson ratio in coalescence is controlled by a parameter $r_{\mathrm{BM}}^{\mathrm{HQ}}$, analogous in spirit to the $r_{\mathrm{BM}}$ parameter introduced in the improved quark coalescence in Ref.~\cite{He:2017tla,Zhang:2025pqu}.
A larger value of $r_{\mathrm{BM}}^{\mathrm{HQ}}$ effectively enhances the probability for a charm quark to form a baryon rather than a meson.
We find that varying $r_{\mathrm{BM}}^{\mathrm{HQ}}$ has a noticeable effect on the heavy flavor baryon-to-meson ratio in AA, while the effect for pp collisions is relatively small, and this is due to the coalescence probability is smaller in pp compared to that in AA collisions.
In practice, we fix $r_{\mathrm{BM}}^{\mathrm{HQ}}=2$ for charm, which allows us to reproduce the integrated $D^{0}$ yield in central Au+Au collisions without spoiling the description of the pp baseline.

\section{Results and Discussion}\label{sec.III}

\begin{figure*}[htbp]
    \centering
    \includegraphics[width=0.85\textwidth]{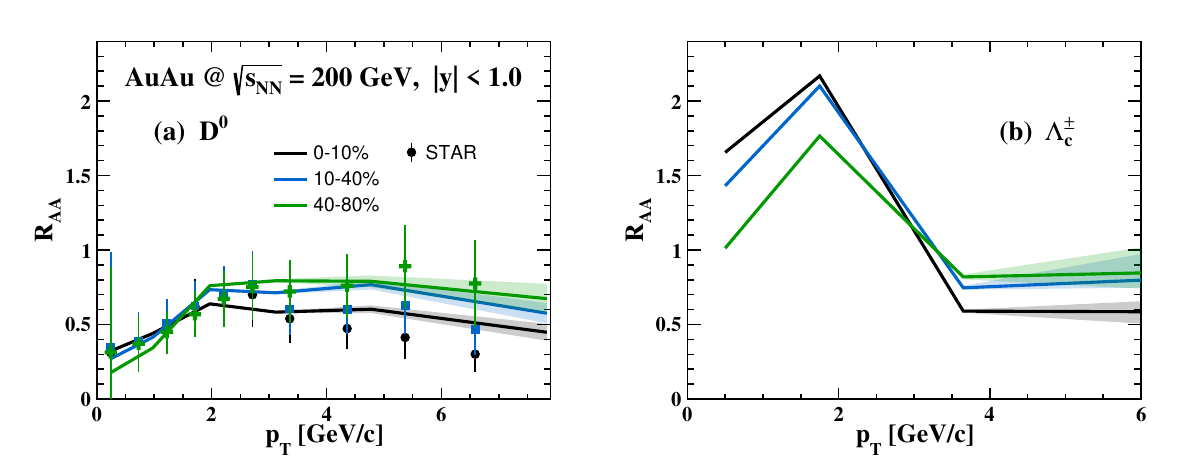}
    \caption{Nuclear modification factors $R_{\mathrm{AA}}$ of (a) $D^{0}$ mesons and (b) $\Lambda_{c}^{+}$ baryons at mid-rapidity in Au+Au collisions at $\sqrt{s_{\text{NN}}}$ = 200 GeV for three centrality intervals. The bands denote the results from the updated string-melting AMPT model, while the markers in panel (a) show the STAR data for $D^{0}$ mesons. For $\Lambda_{c}^{+}$, no experimental $R_{\mathrm{AA}}$ measurements are currently available at this energy. The average number of binary nucleon–nucleon collisions $\langle N_{\mathrm{coll}}\rangle$ for each centrality bin are taken from the corresponding experimental Glauber analysis referenced in the txet.}
    \label{D0_lc_RAA}
\end{figure*}

We now apply the AMPT-SM model to investigate the open charm production in $pp$ and Au+Au collisions at $\sqrt{s_{\mathrm{NN}}} = 200$~GeV. 
In this section we confront the model with existing $D^{0}$ and $\Lambda_{c}^{+}$ measurements from STAR and discuss the implications for charm transport and hadronization in the QCD medium. 
Unless otherwise specified, all results refer to the average over particles and antiparticles.

Figure.~\ref{D0_lc_ptspectra} shows the transverse momentum spectra of $D^{0}$ meson and $\Lambda_{c}^{+}$ baryon at mid-rapidity in $pp$ and in Au+Au collisions for different centrality classes, compared with the STAR data \cite{STAR:2018zdy,STAR:2019ank}. 
As expected, the $p_{\mathrm{T}}$ spectra of charm hadrons exhibit a steep fall with increasing transverse momentum. They are enhanced in central compared to peripheral Au+Au collisions, reflecting the growing number of binary nucleon-nucleon collisions. 
In the case of $D^{0}$ mesons, the AMPT model successfully reproduces the measured $p_{\mathrm{T}}$ spectra in $pp$ and Au+Au collisions within the experimental uncertainties over a broad $p_{\mathrm{T}}$ range. The good description in pp collisions provides a necessary baseline for constructing the nuclear modification factor discussed later.
In Au+Au collisions, the model’s accurate description across centrality classes indicates that the improved AMPT version captures the essential features of medium effects on open charm hadron production at RHIC energies.
Regarding $\Lambda_{c}^{+}$ production, the currently available data are limited to the 10–80\% centrality interval in Au+Au collisions. Within experimental uncertainties, the AMPT model calculations in this centrality range are consistent with the measurements.
We also made predictions for $\Lambda_{c}^{+}$ productions in other centralities of Au+Au collisions and in $pp$ collisions, as shown by Fig.~\ref{D0_lc_ptspectra}, which will be used to construct the nuclear modification factor of $\Lambda_{c}^{+}$ baryon and $\Lambda_{c}^{+}/D^{0}$ ratios in kinematic regions where data are not yet available.

The medium effects on open charm production are often quantified by the nuclear modification factor:

\begin{equation}
  R_{\mathrm{AA}}(p_{\mathrm{T}}) 
  = \frac{1}{\langle N_{\mathrm{coll}}\rangle}\,
    \frac{\mathrm{d}^{2}N^{AA}/\mathrm{d}p_{\mathrm{T}}\mathrm{d}y}
         {\mathrm{d}^{2}N^{pp}/\mathrm{d}p_{\mathrm{T}}\mathrm{d}y},
  \label{eq:RAA_def}
\end{equation}

where $\langle N_{\mathrm{coll}}\rangle$ is the averaged number of binary nucleon-nucleon collisions for a given centrality class, taken from the Glauber model used in the STAR measurements \cite{STAR:2014wif}. 
In both cases the $pp$ reference in the denominator is from the AMPT calculation. For $D^{0}$, the model reproduces the existing $pp$ data at $\sqrt{s_{\mathrm{NN}}}=200$~GeV, as shown in Fig.~\ref{D0_lc_ptspectra}, while for $\Lambda_{c}^{+}$ it provides a prediction in the absence of data at this energy.

Figure.~\ref{D0_lc_RAA} shows the $R_{\mathrm{AA}}(p_{\mathrm{T}})$ for $D^{0}$ mesons and $\Lambda_{c}^{+}$ baryons at mid-rapidity in Au+Au collisions at $\sqrt{s_{\mathrm{NN}}}=200$~GeV.
For all centrality intervals, the $R_{\mathrm{AA}}$ of $D^{0}$ rises from values below unity at very low $p_{\mathrm{T}}$ to a maximum near $p_{\mathrm{T}}\sim 2$-3~GeV$/c$, and then decreases gradually at higher $p_{\mathrm{T}}$, where a sizable suppression with respect to the binary-scaled $pp$ reference is observed.
At lower $p_{\mathrm{T}}$, the enhancing behavior of the $R_{\mathrm{AA}}$ dependent on $p_{\mathrm{T}}$ can be related to the interplay of the initial Cronin momentum broadening effect and the in-medium thermalization dynamics,  such as momentum redistribution via drag forces and enhanced hadronization through coalescence.
At higher $p_{\mathrm{T}}$, energy loss in the deconfined medium dominates, and the spectrum is pushed to lower transverse momenta, resulting in $R_{\mathrm{AA}}<$ 1.
The magnitude of the high-$p_{\mathrm{T}}$ suppression increases from peripheral to central collisions, which is consistent with the expectation that charm quarks traverse a denser and longer-lived medium and therefore suffer stronger energy loss in more central events. 
Since the $pp$ baseline is reasonably reproduced by the AMPT model, the resulting $R_{\mathrm{AA}}$ provides a reliable observation of medium-induced modifications to charm quark. The observed centrality and $p_{\mathrm{T}}$ dependence of the $D^{0}$ $R_{\mathrm{AA}}$ indicates that charm quarks experience substantial interactions in the partonic medium prior to the hadronization stage.

For $\Lambda_{c}^{+}$, the $R_{\mathrm{AA}}$ in Fig.~\ref{D0_lc_RAA}(b) represents a model prediction. 
Compared with $D^{0}$, the calculated $\Lambda_{c}^{+}$ shows a more pronounced enhancement at intermediate $p_{\mathrm{T}}$.
This behavior arises from the increased probability for charm quarks to hadronize into baryons via coalescence in the nuclear environment, as encoded by the coalescence criteria $(p_{\mathrm{r}},p_{\mathrm{m}})$ and the parameter $r_{\mathrm{BM}}^{\mathrm{HQ}}$.
At higher $p_{\mathrm{T}}$, where independent fragmentation becomes more important, the $\Lambda_{c}^{+}$ and $D^{0}$ nuclear modification factors tend to approach each other, indicating that both hadrons inherit comparable energy loss from their parent charm quarks in the fragmentation dominated kinematic regime. 
Overall, the predicted $R_{\mathrm{AA}}$ of $\Lambda_{c}^{+}$ exhibits a transition from coalescence-dominated enhancement at low to intermediate $p_{\mathrm{T}}$ to a fragmentation-dominated regime at higher $p_{\mathrm{T}}$, consistent with the behavior observed for $D^{0}$ within the same framework.

\begin{figure}[htbp]
    \centering
    \includegraphics[width=0.45\textwidth]{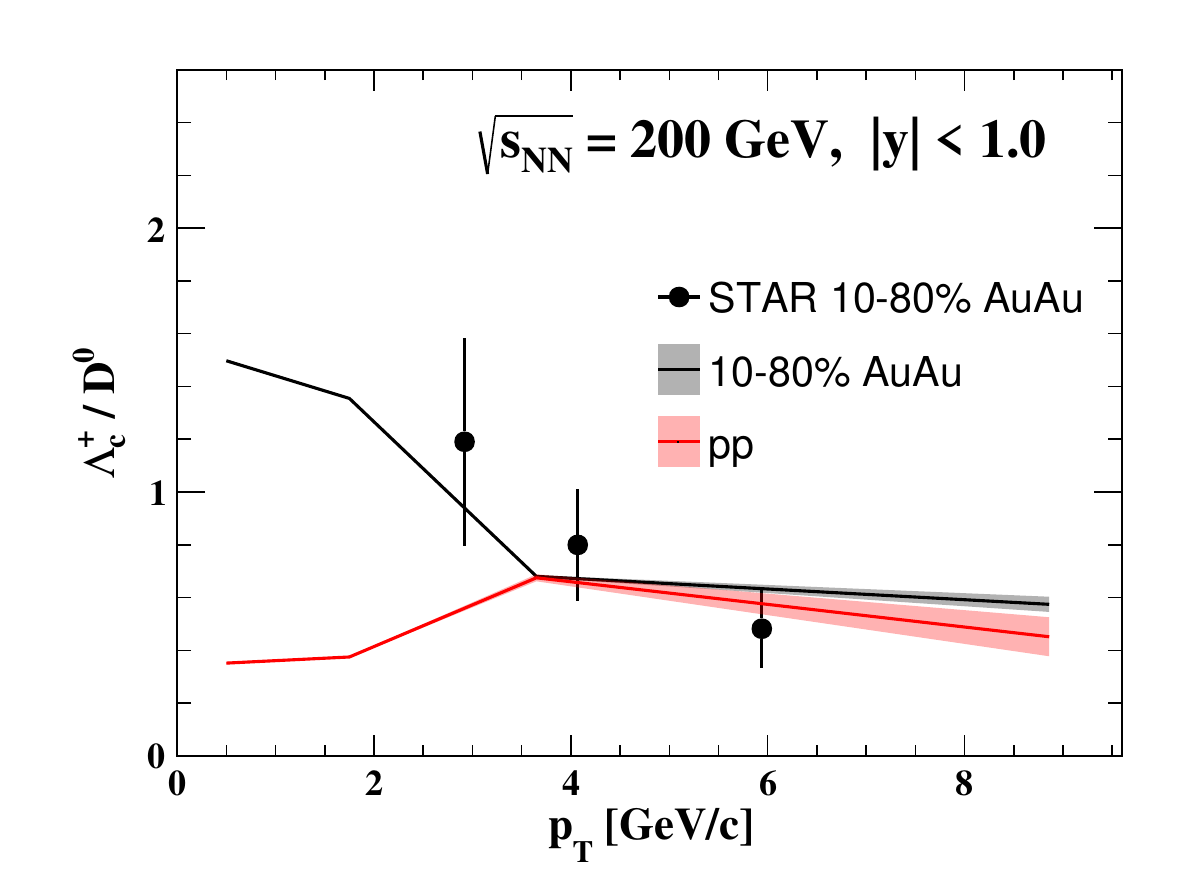}
    \caption{Transverse momentum dependence of the $\Lambda_{c}^{+}/D^{0}$ yield ratio at mid-rapidity in $pp$ and 10-80\% Au+Au collisions at $\sqrt{s_{\mathrm{NN}}} = 200$~GeV.
    Curves show results from the updated string-melting AMPT model, while the data points denote the STAR measurement~\cite{STAR:2018zdy}.}
    \label{LambdactoD0_pp_AA}
\end{figure}

The ratio of charm-baryon to charm-meson yields has emerged as a particularly sensitive probe of different hadronization mechanisms, complementary to the information in $R_{\mathrm{AA}}$.
Figure~\ref{LambdactoD0_pp_AA} presents the $\Lambda_{c}^{+}/D^{0}$ ratio as a function of $p_{\mathrm{T}}$ in $pp$ and in 10-80\% Au+Au collisions at $\sqrt{s_{\mathrm{NN}}} = 200$~GeV.
The Au+Au result is compared with the STAR data~\cite{STAR:2018zdy}, while the $pp$ curve represents a model prediction to compare with Au+Au.
For 10-80\% Au+Au collisions, the AMPT calculation reproduces both the magnitude and the $p_{\mathrm{T}}$ dependence of the data within the uncertainties, yielding $\Lambda_{c}^{+}/D^{0}$ values of order $0.5$-$0.8$ in the range $3 \lesssim p_{\mathrm{T}} \lesssim 6$~GeV$/c$.
The observed enhancement of the ratio in Au+Au collisions at low to intermediate $p_{\mathrm{T}}$ relative to pp collisions supports the picture where charm quarks hadronize via coalescence with thermal light quarks in the dense medium. This process, implemented via the coalescence criteria in Eq.~\ref{eq:coal_pr} and Eq.~\ref{eq:coal_pm} with parameter $r_{\mathrm{BM}}^{\mathrm{HQ}}$, naturally enhances baryon formation relative to vacuum-like fragmentation.
In contrast, the predicted $\Lambda_{c}^{+}/D^{0}$ ratio in $pp$ collisions remains systematically below the Au+Au result at low to intermediate $p_{\mathrm{T}}$, reflecting the reduced space-time volume and lower parton density in $pp$ events, where charm hadronization is closer to vacuum-like fragmentation.
The AMPT framework, by implementing the combined hadronization mechanisms for charm quarks considering both coalescence and independent fragmentation, provides a unified description of this system-dependent behavior.

\begin{figure}[!htb]
    \centering
    \includegraphics[width=0.45\textwidth]{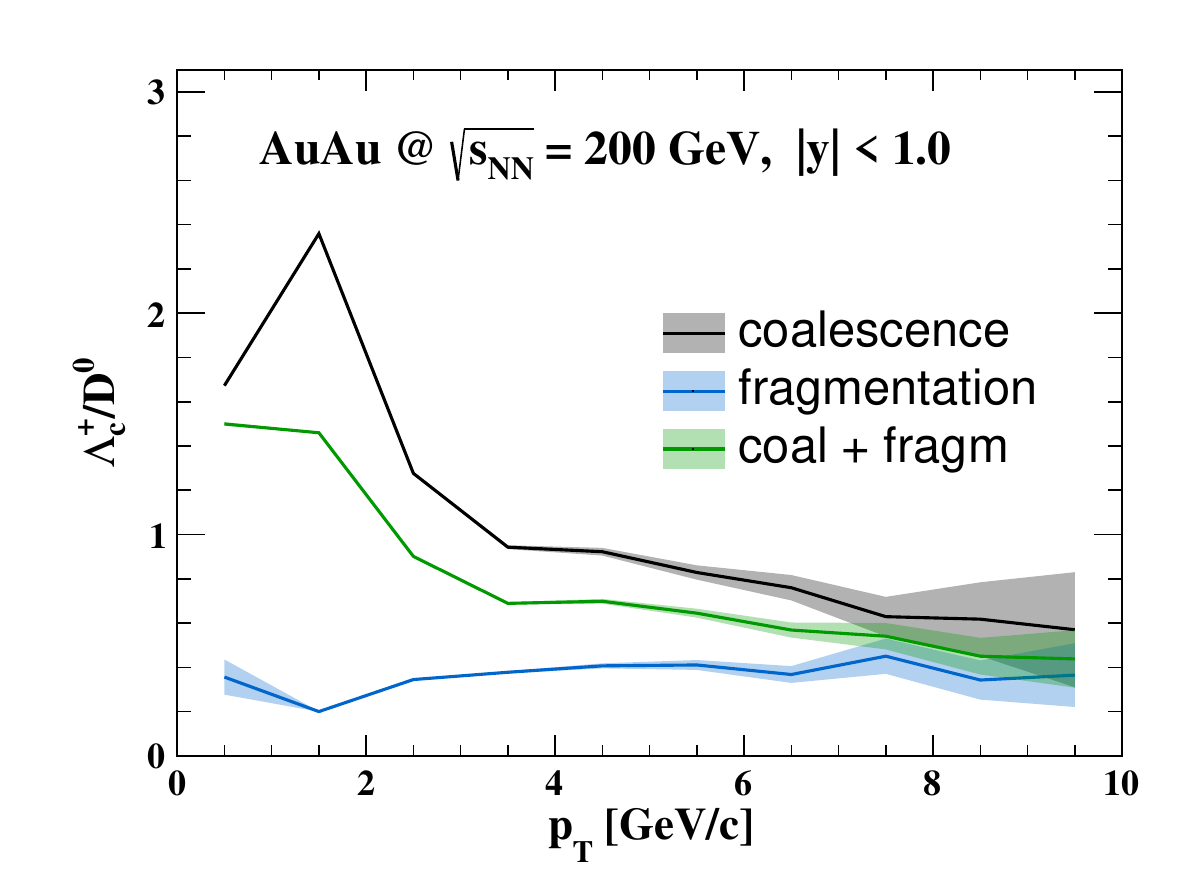}
    \caption{Transverse momentum dependence of the $\Lambda_{c}^{+}/D^{0}$ yield ratio at mid-rapidity in Au+Au collisions at $\sqrt{s_{\mathrm{NN}}} = 200$~GeV from the updated AMPT model, separated into contributions from charm-quark coalescence, independent fragmentation, and their mix.}
    \label{LambdactoD0_AuAu_coal_frag}
\end{figure}

To further elucidate the role of the two hadronization mechanisms in shaping the charm baryon to meson ratio, Fig.~\ref{LambdactoD0_AuAu_coal_frag} breaks down the $p_{\mathrm{T}}$ dependent $\Lambda_{c}^{+}/D^{0}$ ratio in Au+Au collisions at $\sqrt{s_{\mathrm{NN}}} = 200$~GeV into contributions from pure coalescence, pure independent fragmentation, and their combined result in the full model, directly elucidating their roles in charm baryon production.

The coalescence component alone generates a large enhancement of $\Lambda_{c}^{+}/D^{0}$ at low and intermediate transverse momenta, with values well above unity for $p_{\mathrm{T}}\lesssim 3$~GeV$/c$. 
This behavior reflects the increased probability for a charm quark to recombine with nearby light quarks in phase space, which is further enhanced by the enlarged effective coalescence window in Au+Au collisions.
In addition to the high densities of thermal light quarks and their strong collective flow at low and intermediate $p_{\mathrm{T}}$, the use of a larger $m_{\mathrm{T}}$ selection provides a coalescence condition complementary to the constraints on relative distance and invariant mass, allowing baryon formation once either criterion is satisfied.
The combined effect leads to a pronounced enhancement of the $\Lambda_c^+/D^0$ ratio at low and intermediate transverse momenta of pure coalescence.
The pure independent fragmentation component yields a substantially smaller, nearly flat $\Lambda_{c}^{+}/D^{0}$ ratio. This behavior is characteristic of vacuum-like, meson-dominated charm production, where the baryon fraction is determined by suppressed weights in the fragmentation functions rather than the dynamic partonic environment.

When both coalescence and fragmentation are included, the resulting $\Lambda_{c}^{+}/D^{0}$ ratio lies between the two limiting cases. 
At low and intermediate $p_{\mathrm{T}}$, coalescence dominates and drives the enhancement of $\Lambda_{c}^{+}/D^{0}$ observed in Au+Au data relative to the $pp$ baseline. At higher $p_{\mathrm{T}}$, the relative weight of fragmentation increases and drives the charm baryon to meson ratio downward. 
This pattern is consistent with the qualitative picture outlined in the introduction, that charm hadronization in heavy-ion collisions is governed by a momentum-dependent competition between medium-induced coalescence and vacuum-like fragmentation, with coalescence playing a leading role in the low- and intermediate-$p_{\mathrm{T}}$ region and fragmentation dominating at high $p_{\mathrm{T}}$.

Before concluding, we comment on the model systematics and the scope of the present study. The results presented here are obtained within a unified modeling framework for charm transport and hadronization, where the same set of physical ingredients is applied to both $pp$ and Au+Au collisions, allowing for system-dependent comparisons. In particular, different transverse mass selections are employed in small and large systems to account for their distinct partonic environments and to regulate the relative importance of coalescence and fragmentation.

Within this setup, the model provides a simultaneous description of the measured $D^{0}$ spectra in $pp$ and Au+Au collisions, as well as the $\Lambda_{c}^{+}$ spectrum and $\Lambda_{c}^{+}/D^{0}$ ratio in 10–80\% Au+Au collisions within uncertainties. The overall level of agreement is driven by the interplay between the effective coalescence, the heavy flavor baryon-to-meson ratio $r_{\mathrm{BM}}^{\mathrm{HQ}}$, and the transverse-mass–based selection that enhances low-$p_{\mathrm{T}}$ coalescence in large systems, together with the inclusion of an independent fragmentation channel that dominates at higher $p_{\mathrm{T}}$.

At the same time, it should be emphasized that the present agreement does not uniquely constrain the hadronization parameters; however, the qualitative conclusion that the enhancement of ($\Lambda_c/D^0$) at low and intermediate ($p_{\mathrm{T}}$) is driven by charm-light quark coalescence remains robust against reasonable variations of these parameters. A quantitative assessment of the remaining model uncertainties would require a systematic variation of the coalescence criteria, the baryon-to-meson ratio, and the transverse mass selection, as well as a detailed exploration of their correlations. Such an extensive scan of the parameter space, while highly desirable, is beyond the scope of the present work.

\section{Summary}\label{sec.IV}

In summary, we have employed an updated string-melting version of the AMPT model to investigate open charm production in $pp$ and Au+Au collisions at $\sqrt{s_{\mathrm{NN}}} = 200$~GeV. 
Several ingredients have been implemented specifically for the heavy flavor sector. 
Charm-anticharm pairs are taken directly from the HIJING initial conditions and propagated with a finite formation time before entering the partonic cascade without going through the string melting procedure. An initial-state Cronin momentum broadening is applied to the extracted $Q\bar{Q}$ pairs, and charm quarks hadronize through a hybrid hadronization framework allowing a competitive interplay between quark coalescence and independent fragmentation.

Within this framework, the updated AMPT model provides a reasonable description of the measured $D^{0}$ transverse momentum spectra in both $pp$ and Au+Au collisions, as well as the available $\Lambda_{c}^{+}$ spectra in Au+Au at 10-80\% centrality. Thus, the model can reproduce the main qualitative features of the $D^{0}$ nuclear modification factor and the $\Lambda_{c}^{+}/D^{0}$ ratio in 10–80\% Au+Au collisions. Our results demonstrate that the observed baryon enhancement at intermediate $p_{\mathrm{T}}$ is driven by the coalescence mechanism, while the high-$p_{\mathrm{T}}$ region is governed by fragmentation. This finding reinforces the coalescence paradigm established in the light-flavor sector and suggests that heavy quarks thermalize partially with the medium and recombine with light quarks in the collective partonic expansion before hadronization. These findings demonstrate the importance of a unified microscopic description of charm dynamics that simultaneously accounts for the space-time evolution of the medium and the momentum-dependent competition between coalescence and fragmentation within a single transport framework.

This unified framework built in the current study offers a robust platform for exploring charm dynamics across a broader range of nuclear environments.
Future measurements of the $\Lambda_c^+$ nuclear modification factor and its centrality dependence at RHIC energies will provide a direct test of the coalescence-dominated hadronization scenario proposed here.

{\bf Acknowledgments:}
We are grateful for discussions with Prof. Zi-Wei Lin.
This work is supported in part by the National Key Research and Development Program of China under Contract Nos. 2024YFA1610700 and 2022YFA1604900 (S.S); National Natural Science Foundation of China 12405159(C.Z), Natural Science Foundation of Hubei Province (2024AFB136) (C.Z); Key Laboratory of Quark and Lepton Physics QLPL2023P01(C.Z) and QLPL2025P01 (L.Z).
The numerical calculations were performed on a GPU cluster located at the Nuclear Science Computing Center, Central China Normal University (NSC3).

\bibliography{example} 

\begin{thebibliography}{59}%
\makeatletter
\providecommand \@ifxundefined [1]{%
 \@ifx{#1\undefined}
}%
\providecommand \@ifnum [1]{%
 \ifnum #1\expandafter \@firstoftwo
 \else \expandafter \@secondoftwo
 \fi
}%
\providecommand \@ifx [1]{%
 \ifx #1\expandafter \@firstoftwo
 \else \expandafter \@secondoftwo
 \fi
}%
\providecommand \natexlab [1]{#1}%
\providecommand \enquote  [1]{``#1''}%
\providecommand \bibnamefont  [1]{#1}%
\providecommand \bibfnamefont [1]{#1}%
\providecommand \citenamefont [1]{#1}%
\providecommand \href@noop [0]{\@secondoftwo}%
\providecommand \href [0]{\begingroup \@sanitize@url \@href}%
\providecommand \@href[1]{\@@startlink{#1}\@@href}%
\providecommand \@@href[1]{\endgroup#1\@@endlink}%
\providecommand \@sanitize@url [0]{\catcode `\\12\catcode `\$12\catcode
  `\&12\catcode `\#12\catcode `\^12\catcode `\_12\catcode `\%12\relax}%
\providecommand \@@startlink[1]{}%
\providecommand \@@endlink[0]{}%
\providecommand \url  [0]{\begingroup\@sanitize@url \@url }%
\providecommand \@url [1]{\endgroup\@href {#1}{\urlprefix }}%
\providecommand \urlprefix  [0]{URL }%
\providecommand \Eprint [0]{\href }%
\providecommand \doibase [0]{https://doi.org/}%
\providecommand \selectlanguage [0]{\@gobble}%
\providecommand \bibinfo  [0]{\@secondoftwo}%
\providecommand \bibfield  [0]{\@secondoftwo}%
\providecommand \translation [1]{[#1]}%
\providecommand \BibitemOpen [0]{}%
\providecommand \bibitemStop [0]{}%
\providecommand \bibitemNoStop [0]{.\EOS\space}%
\providecommand \EOS [0]{\spacefactor3000\relax}%
\providecommand \BibitemShut  [1]{\csname bibitem#1\endcsname}%
\let\auto@bib@innerbib\@empty
\bibitem [{\citenamefont {Gyulassy}\ and\ \citenamefont
  {McLerran}(2005)}]{Gyulassy:2004zy}%
  \BibitemOpen
  \bibfield  {author} {\bibinfo {author} {\bibfnamefont {M.}~\bibnamefont
  {Gyulassy}}\ and\ \bibinfo {author} {\bibfnamefont {L.}~\bibnamefont
  {McLerran}},\ }\bibfield  {title} {\bibinfo {title} {{New forms of QCD matter
  discovered at RHIC}},\ }\href
  {https://doi.org/10.1016/j.nuclphysa.2004.10.034} {\bibfield  {journal}
  {\bibinfo  {journal} {Nucl. Phys. A}\ }\textbf {\bibinfo {volume} {750}},\
  \bibinfo {pages} {30} (\bibinfo {year} {2005})},\ \Eprint
  {https://arxiv.org/abs/nucl-th/0405013} {arXiv:nucl-th/0405013} \BibitemShut
  {NoStop}%
\bibitem [{\citenamefont {Adcox}\ \emph {et~al.}(2005)\citenamefont {Adcox}
  \emph {et~al.}}]{PHENIX:2004vcz}%
  \BibitemOpen
  \bibfield  {author} {\bibinfo {author} {\bibfnamefont {K.}~\bibnamefont
  {Adcox}} \emph {et~al.} (\bibinfo {collaboration} {PHENIX}),\ }\bibfield
  {title} {\bibinfo {title} {{Formation of dense partonic matter in
  relativistic nucleus-nucleus collisions at RHIC: Experimental evaluation by
  the PHENIX collaboration}},\ }\href
  {https://doi.org/10.1016/j.nuclphysa.2005.03.086} {\bibfield  {journal}
  {\bibinfo  {journal} {Nucl. Phys. A}\ }\textbf {\bibinfo {volume} {757}},\
  \bibinfo {pages} {184} (\bibinfo {year} {2005})},\ \Eprint
  {https://arxiv.org/abs/nucl-ex/0410003} {arXiv:nucl-ex/0410003} \BibitemShut
  {NoStop}%
\bibitem [{\citenamefont {Adams}\ \emph
  {et~al.}(2005{\natexlab{a}})\citenamefont {Adams} \emph
  {et~al.}}]{STAR:2005gfr}%
  \BibitemOpen
  \bibfield  {author} {\bibinfo {author} {\bibfnamefont {J.}~\bibnamefont
  {Adams}} \emph {et~al.} (\bibinfo {collaboration} {STAR}),\ }\bibfield
  {title} {\bibinfo {title} {{Experimental and theoretical challenges in the
  search for the quark gluon plasma: The STAR Collaboration's critical
  assessment of the evidence from RHIC collisions}},\ }\href
  {https://doi.org/10.1016/j.nuclphysa.2005.03.085} {\bibfield  {journal}
  {\bibinfo  {journal} {Nucl. Phys. A}\ }\textbf {\bibinfo {volume} {757}},\
  \bibinfo {pages} {102} (\bibinfo {year} {2005}{\natexlab{a}})},\ \Eprint
  {https://arxiv.org/abs/nucl-ex/0501009} {arXiv:nucl-ex/0501009} \BibitemShut
  {NoStop}%
\bibitem [{\citenamefont {Chen}\ \emph {et~al.}(2024)\citenamefont {Chen} \emph
  {et~al.}}]{Chen:2024aom}%
  \BibitemOpen
  \bibfield  {author} {\bibinfo {author} {\bibfnamefont {J.}~\bibnamefont
  {Chen}} \emph {et~al.},\ }\bibfield  {title} {\bibinfo {title} {{Properties
  of the QCD matter: review of selected results from the relativistic heavy ion
  collider beam energy scan (RHIC BES) program}},\ }\href
  {https://doi.org/10.1007/s41365-024-01591-2} {\bibfield  {journal} {\bibinfo
  {journal} {Nucl. Sci. Tech.}\ }\textbf {\bibinfo {volume} {35}},\ \bibinfo
  {pages} {214} (\bibinfo {year} {2024})},\ \Eprint
  {https://arxiv.org/abs/2407.02935} {arXiv:2407.02935 [nucl-ex]} \BibitemShut
  {NoStop}%
\bibitem [{\citenamefont {Luo}\ \emph {et~al.}(2020)\citenamefont {Luo},
  \citenamefont {Shi}, \citenamefont {Xu},\ and\ \citenamefont
  {Zhang}}]{Luo:2020pef}%
  \BibitemOpen
  \bibfield  {author} {\bibinfo {author} {\bibfnamefont {X.}~\bibnamefont
  {Luo}}, \bibinfo {author} {\bibfnamefont {S.}~\bibnamefont {Shi}}, \bibinfo
  {author} {\bibfnamefont {N.}~\bibnamefont {Xu}},\ and\ \bibinfo {author}
  {\bibfnamefont {Y.}~\bibnamefont {Zhang}},\ }\bibfield  {title} {\bibinfo
  {title} {{A Study of the Properties of the QCD Phase Diagram in High-Energy
  Nuclear Collisions}},\ }\href {https://doi.org/10.3390/particles3020022}
  {\bibfield  {journal} {\bibinfo  {journal} {Particles}\ }\textbf {\bibinfo
  {volume} {3}},\ \bibinfo {pages} {278} (\bibinfo {year} {2020})},\ \Eprint
  {https://arxiv.org/abs/2004.00789} {arXiv:2004.00789 [nucl-ex]} \BibitemShut
  {NoStop}%
\bibitem [{\citenamefont {Bzdak}\ \emph {et~al.}(2020)\citenamefont {Bzdak}
  \emph {et~al.}}]{Bzdak:2019pkr}%
  \BibitemOpen
  \bibfield  {author} {\bibinfo {author} {\bibfnamefont {A.}~\bibnamefont
  {Bzdak}} \emph {et~al.},\ }\bibfield  {title} {\bibinfo {title} {{Mapping the
  Phases of Quantum Chromodynamics with Beam Energy Scan}},\ }\href
  {https://doi.org/10.1016/j.physrep.2020.01.005} {\bibfield  {journal}
  {\bibinfo  {journal} {Phys. Rept.}\ }\textbf {\bibinfo {volume} {853}},\
  \bibinfo {pages} {1} (\bibinfo {year} {2020})}\BibitemShut {NoStop}%
\bibitem [{\citenamefont {Muller}\ and\ \citenamefont
  {Wang}(1992)}]{Muller:1992xn}%
  \BibitemOpen
  \bibfield  {author} {\bibinfo {author} {\bibfnamefont {B.}~\bibnamefont
  {Muller}}\ and\ \bibinfo {author} {\bibfnamefont {X.-N.}\ \bibnamefont
  {Wang}},\ }\bibfield  {title} {\bibinfo {title} {{Probing parton
  thermalization time with charm production}},\ }\href
  {https://doi.org/10.1103/PhysRevLett.68.2437} {\bibfield  {journal} {\bibinfo
   {journal} {Phys. Rev. Lett.}\ }\textbf {\bibinfo {volume} {68}},\ \bibinfo
  {pages} {2437} (\bibinfo {year} {1992})}\BibitemShut {NoStop}%
\bibitem [{\citenamefont {Lin}\ and\ \citenamefont
  {Gyulassy}(1995)}]{Lin:1994xma}%
  \BibitemOpen
  \bibfield  {author} {\bibinfo {author} {\bibfnamefont {Z.-w.}\ \bibnamefont
  {Lin}}\ and\ \bibinfo {author} {\bibfnamefont {M.}~\bibnamefont {Gyulassy}},\
  }\bibfield  {title} {\bibinfo {title} {{Open charm as a probe of
  preequilibrium dynamics in nuclear collisions}},\ }\href
  {https://doi.org/10.1103/PhysRevC.52.440} {\bibfield  {journal} {\bibinfo
  {journal} {Phys. Rev. C}\ }\textbf {\bibinfo {volume} {51}},\ \bibinfo
  {pages} {2177} (\bibinfo {year} {1995})},\ \bibinfo {note} {[Erratum:
  Phys.Rev.C 52, 440 (1995)]},\ \Eprint {https://arxiv.org/abs/nucl-th/9409007}
  {arXiv:nucl-th/9409007} \BibitemShut {NoStop}%
\bibitem [{\citenamefont {van Hees}\ \emph {et~al.}(2006)\citenamefont {van
  Hees}, \citenamefont {Greco},\ and\ \citenamefont {Rapp}}]{vanHees:2005wb}%
  \BibitemOpen
  \bibfield  {author} {\bibinfo {author} {\bibfnamefont {H.}~\bibnamefont {van
  Hees}}, \bibinfo {author} {\bibfnamefont {V.}~\bibnamefont {Greco}},\ and\
  \bibinfo {author} {\bibfnamefont {R.}~\bibnamefont {Rapp}},\ }\bibfield
  {title} {\bibinfo {title} {{Heavy-quark probes of the quark-gluon plasma at
  RHIC}},\ }\href {https://doi.org/10.1103/PhysRevC.73.034913} {\bibfield
  {journal} {\bibinfo  {journal} {Phys. Rev. C}\ }\textbf {\bibinfo {volume}
  {73}},\ \bibinfo {pages} {034913} (\bibinfo {year} {2006})},\ \Eprint
  {https://arxiv.org/abs/nucl-th/0508055} {arXiv:nucl-th/0508055} \BibitemShut
  {NoStop}%
\bibitem [{\citenamefont {Brambilla}\ \emph {et~al.}(2011)\citenamefont
  {Brambilla} \emph {et~al.}}]{Brambilla:2010cs}%
  \BibitemOpen
  \bibfield  {author} {\bibinfo {author} {\bibfnamefont {N.}~\bibnamefont
  {Brambilla}} \emph {et~al.},\ }\bibfield  {title} {\bibinfo {title} {{Heavy
  Quarkonium: Progress, Puzzles, and Opportunities}},\ }\href
  {https://doi.org/10.1140/epjc/s10052-010-1534-9} {\bibfield  {journal}
  {\bibinfo  {journal} {Eur. Phys. J. C}\ }\textbf {\bibinfo {volume} {71}},\
  \bibinfo {pages} {1534} (\bibinfo {year} {2011})},\ \Eprint
  {https://arxiv.org/abs/1010.5827} {arXiv:1010.5827 [hep-ph]} \BibitemShut
  {NoStop}%
\bibitem [{\citenamefont {Andronic}\ \emph {et~al.}(2016)\citenamefont
  {Andronic} \emph {et~al.}}]{Andronic:2015wma}%
  \BibitemOpen
  \bibfield  {author} {\bibinfo {author} {\bibfnamefont {A.}~\bibnamefont
  {Andronic}} \emph {et~al.},\ }\bibfield  {title} {\bibinfo {title}
  {{Heavy-flavour and quarkonium production in the LHC era: from
  proton{\textendash}proton to heavy-ion collisions}},\ }\href
  {https://doi.org/10.1140/epjc/s10052-015-3819-5} {\bibfield  {journal}
  {\bibinfo  {journal} {Eur. Phys. J. C}\ }\textbf {\bibinfo {volume} {76}},\
  \bibinfo {pages} {107} (\bibinfo {year} {2016})},\ \Eprint
  {https://arxiv.org/abs/1506.03981} {arXiv:1506.03981 [nucl-ex]} \BibitemShut
  {NoStop}%
\bibitem [{\citenamefont {Adamczyk}\ \emph {et~al.}(2012)\citenamefont
  {Adamczyk} \emph {et~al.}}]{STAR:2012nbd}%
  \BibitemOpen
  \bibfield  {author} {\bibinfo {author} {\bibfnamefont {L.}~\bibnamefont
  {Adamczyk}} \emph {et~al.} (\bibinfo {collaboration} {STAR}),\ }\bibfield
  {title} {\bibinfo {title} {{Measurements of $D^{0}$ and $D^{*}$ Production in
  $p+p$ Collisions at $\sqrt{s} = 200$ GeV}},\ }\href
  {https://doi.org/10.1103/PhysRevD.86.072013} {\bibfield  {journal} {\bibinfo
  {journal} {Phys. Rev. D}\ }\textbf {\bibinfo {volume} {86}},\ \bibinfo
  {pages} {072013} (\bibinfo {year} {2012})},\ \Eprint
  {https://arxiv.org/abs/1204.4244} {arXiv:1204.4244 [nucl-ex]} \BibitemShut
  {NoStop}%
\bibitem [{\citenamefont {Adamczyk}\ \emph {et~al.}(2014)\citenamefont
  {Adamczyk} \emph {et~al.}}]{STAR:2014wif}%
  \BibitemOpen
  \bibfield  {author} {\bibinfo {author} {\bibfnamefont {L.}~\bibnamefont
  {Adamczyk}} \emph {et~al.} (\bibinfo {collaboration} {STAR}),\ }\bibfield
  {title} {\bibinfo {title} {{Observation of $D^0$ Meson Nuclear Modifications
  in Au+Au Collisions at $\sqrt{s_{NN}}=200$ GeV}},\ }\href
  {https://doi.org/10.1103/PhysRevLett.113.142301} {\bibfield  {journal}
  {\bibinfo  {journal} {Phys. Rev. Lett.}\ }\textbf {\bibinfo {volume} {113}},\
  \bibinfo {pages} {142301} (\bibinfo {year} {2014})},\ \bibinfo {note}
  {[Erratum: Phys.Rev.Lett. 121, 229901 (2018)]},\ \Eprint
  {https://arxiv.org/abs/1404.6185} {arXiv:1404.6185 [nucl-ex]} \BibitemShut
  {NoStop}%
\bibitem [{\citenamefont {Adam}\ \emph {et~al.}(2019)\citenamefont {Adam} \emph
  {et~al.}}]{STAR:2018zdy}%
  \BibitemOpen
  \bibfield  {author} {\bibinfo {author} {\bibfnamefont {J.}~\bibnamefont
  {Adam}} \emph {et~al.} (\bibinfo {collaboration} {STAR}),\ }\bibfield
  {title} {\bibinfo {title} {{Centrality and transverse momentum dependence of
  $D^0$-meson production at mid-rapidity in Au+Au collisions at ${\sqrt{s_{\rm
  NN}} = \rm{200\,GeV}}$}},\ }\href
  {https://doi.org/10.1103/PhysRevC.99.034908} {\bibfield  {journal} {\bibinfo
  {journal} {Phys. Rev. C}\ }\textbf {\bibinfo {volume} {99}},\ \bibinfo
  {pages} {034908} (\bibinfo {year} {2019})},\ \Eprint
  {https://arxiv.org/abs/1812.10224} {arXiv:1812.10224 [nucl-ex]} \BibitemShut
  {NoStop}%
\bibitem [{\citenamefont {Adam}\ \emph {et~al.}(2020)\citenamefont {Adam} \emph
  {et~al.}}]{STAR:2019ank}%
  \BibitemOpen
  \bibfield  {author} {\bibinfo {author} {\bibfnamefont {J.}~\bibnamefont
  {Adam}} \emph {et~al.} (\bibinfo {collaboration} {STAR}),\ }\bibfield
  {title} {\bibinfo {title} {{First measurement of $\Lambda_c$ baryon
  production in Au+Au collisions at $\sqrt{s_{\rm NN}}$ = 200 GeV}},\ }\href
  {https://doi.org/10.1103/PhysRevLett.124.172301} {\bibfield  {journal}
  {\bibinfo  {journal} {Phys. Rev. Lett.}\ }\textbf {\bibinfo {volume} {124}},\
  \bibinfo {pages} {172301} (\bibinfo {year} {2020})},\ \Eprint
  {https://arxiv.org/abs/1910.14628} {arXiv:1910.14628 [nucl-ex]} \BibitemShut
  {NoStop}%
\bibitem [{\citenamefont {Adams}\ \emph
  {et~al.}(2005{\natexlab{b}})\citenamefont {Adams} \emph
  {et~al.}}]{STAR:2004ocv}%
  \BibitemOpen
  \bibfield  {author} {\bibinfo {author} {\bibfnamefont {J.}~\bibnamefont
  {Adams}} \emph {et~al.} (\bibinfo {collaboration} {STAR}),\ }\bibfield
  {title} {\bibinfo {title} {{Open charm yields in d + Au collisions at
  s(NN)**(1/2) = 200-GeV}},\ }\href
  {https://doi.org/10.1103/PhysRevLett.94.062301} {\bibfield  {journal}
  {\bibinfo  {journal} {Phys. Rev. Lett.}\ }\textbf {\bibinfo {volume} {94}},\
  \bibinfo {pages} {062301} (\bibinfo {year} {2005}{\natexlab{b}})},\ \Eprint
  {https://arxiv.org/abs/nucl-ex/0407006} {arXiv:nucl-ex/0407006} \BibitemShut
  {NoStop}%
\bibitem [{\citenamefont {Adam}\ \emph {et~al.}(2021)\citenamefont {Adam} \emph
  {et~al.}}]{STAR:2021tte}%
  \BibitemOpen
  \bibfield  {author} {\bibinfo {author} {\bibfnamefont {J.}~\bibnamefont
  {Adam}} \emph {et~al.} (\bibinfo {collaboration} {STAR}),\ }\bibfield
  {title} {\bibinfo {title} {{Observation of $D_{s}^{\pm}/D^0$ enhancement in
  Au+Au collisions at $\sqrt{s_{_{NN}}}$ = 200 GeV}},\ }\href
  {https://doi.org/10.1103/PhysRevLett.127.092301} {\bibfield  {journal}
  {\bibinfo  {journal} {Phys. Rev. Lett.}\ }\textbf {\bibinfo {volume} {127}},\
  \bibinfo {pages} {092301} (\bibinfo {year} {2021})},\ \Eprint
  {https://arxiv.org/abs/2101.11793} {arXiv:2101.11793 [hep-ex]} \BibitemShut
  {NoStop}%
\bibitem [{\citenamefont {Adamczyk}\ \emph {et~al.}(2017)\citenamefont
  {Adamczyk} \emph {et~al.}}]{STAR:2017kkh}%
  \BibitemOpen
  \bibfield  {author} {\bibinfo {author} {\bibfnamefont {L.}~\bibnamefont
  {Adamczyk}} \emph {et~al.} (\bibinfo {collaboration} {STAR}),\ }\bibfield
  {title} {\bibinfo {title} {{Measurement of $D^0$ Azimuthal Anisotropy at
  Midrapidity in Au+Au Collisions at $\sqrt{s_{NN}}$=200 GeV}},\ }\href
  {https://doi.org/10.1103/PhysRevLett.118.212301} {\bibfield  {journal}
  {\bibinfo  {journal} {Phys. Rev. Lett.}\ }\textbf {\bibinfo {volume} {118}},\
  \bibinfo {pages} {212301} (\bibinfo {year} {2017})},\ \Eprint
  {https://arxiv.org/abs/1701.06060} {arXiv:1701.06060 [nucl-ex]} \BibitemShut
  {NoStop}%
\bibitem [{\citenamefont {Seuster}\ \emph {et~al.}(2006)\citenamefont {Seuster}
  \emph {et~al.}}]{Belle:2005mtx}%
  \BibitemOpen
  \bibfield  {author} {\bibinfo {author} {\bibfnamefont {R.}~\bibnamefont
  {Seuster}} \emph {et~al.} (\bibinfo {collaboration} {Belle}),\ }\bibfield
  {title} {\bibinfo {title} {{Charm hadrons from fragmentation and B decays in
  e+ e- annihilation at s**(1/2) = 10.6-GeV}},\ }\href
  {https://doi.org/10.1103/PhysRevD.73.032002} {\bibfield  {journal} {\bibinfo
  {journal} {Phys. Rev. D}\ }\textbf {\bibinfo {volume} {73}},\ \bibinfo
  {pages} {032002} (\bibinfo {year} {2006})},\ \Eprint
  {https://arxiv.org/abs/hep-ex/0506068} {arXiv:hep-ex/0506068} \BibitemShut
  {NoStop}%
\bibitem [{\citenamefont {Aaij}\ \emph {et~al.}(2013)\citenamefont {Aaij} \emph
  {et~al.}}]{LHCb:2013xam}%
  \BibitemOpen
  \bibfield  {author} {\bibinfo {author} {\bibfnamefont {R.}~\bibnamefont
  {Aaij}} \emph {et~al.} (\bibinfo {collaboration} {LHCb}),\ }\bibfield
  {title} {\bibinfo {title} {{Prompt charm production in pp collisions at
  sqrt(s)=7 TeV}},\ }\href {https://doi.org/10.1016/j.nuclphysb.2013.02.010}
  {\bibfield  {journal} {\bibinfo  {journal} {Nucl. Phys. B}\ }\textbf
  {\bibinfo {volume} {871}},\ \bibinfo {pages} {1} (\bibinfo {year} {2013})},\
  \Eprint {https://arxiv.org/abs/1302.2864} {arXiv:1302.2864 [hep-ex]}
  \BibitemShut {NoStop}%
\bibitem [{\citenamefont {Acharya}\ \emph {et~al.}(2019)\citenamefont {Acharya}
  \emph {et~al.}}]{ALICE:2018hbc}%
  \BibitemOpen
  \bibfield  {author} {\bibinfo {author} {\bibfnamefont {S.}~\bibnamefont
  {Acharya}} \emph {et~al.} (\bibinfo {collaboration} {ALICE}),\ }\bibfield
  {title} {\bibinfo {title} {{$\Lambda_\mathrm{c}^+$ production in Pb-Pb
  collisions at $\sqrt{s_{\rm NN}} = 5.02$ TeV}},\ }\href
  {https://doi.org/10.1016/j.physletb.2019.04.046} {\bibfield  {journal}
  {\bibinfo  {journal} {Phys. Lett. B}\ }\textbf {\bibinfo {volume} {793}},\
  \bibinfo {pages} {212} (\bibinfo {year} {2019})},\ \Eprint
  {https://arxiv.org/abs/1809.10922} {arXiv:1809.10922 [nucl-ex]} \BibitemShut
  {NoStop}%
\bibitem [{\citenamefont {Acharya}\ \emph {et~al.}(2023)\citenamefont {Acharya}
  \emph {et~al.}}]{ALICE:2021bib}%
  \BibitemOpen
  \bibfield  {author} {\bibinfo {author} {\bibfnamefont {S.}~\bibnamefont
  {Acharya}} \emph {et~al.} (\bibinfo {collaboration} {ALICE}),\ }\bibfield
  {title} {\bibinfo {title} {{Constraining hadronization mechanisms with
  {\ensuremath{\Lambda}}c+/D0 production ratios in Pb{\textendash}Pb collisions
  at sNN=5.02 TeV}},\ }\href {https://doi.org/10.1016/j.physletb.2023.137796}
  {\bibfield  {journal} {\bibinfo  {journal} {Phys. Lett. B}\ }\textbf
  {\bibinfo {volume} {839}},\ \bibinfo {pages} {137796} (\bibinfo {year}
  {2023})},\ \Eprint {https://arxiv.org/abs/2112.08156} {arXiv:2112.08156
  [nucl-ex]} \BibitemShut {NoStop}%
\bibitem [{\citenamefont {Mangano}\ \emph {et~al.}(1992)\citenamefont
  {Mangano}, \citenamefont {Nason},\ and\ \citenamefont
  {Ridolfi}}]{Mangano:1991jk}%
  \BibitemOpen
  \bibfield  {author} {\bibinfo {author} {\bibfnamefont {M.~L.}\ \bibnamefont
  {Mangano}}, \bibinfo {author} {\bibfnamefont {P.}~\bibnamefont {Nason}},\
  and\ \bibinfo {author} {\bibfnamefont {G.}~\bibnamefont {Ridolfi}},\
  }\bibfield  {title} {\bibinfo {title} {{Heavy quark correlations in hadron
  collisions at next-to-leading order}},\ }\href
  {https://doi.org/10.1016/0550-3213(92)90435-E} {\bibfield  {journal}
  {\bibinfo  {journal} {Nucl. Phys. B}\ }\textbf {\bibinfo {volume} {373}},\
  \bibinfo {pages} {295} (\bibinfo {year} {1992})}\BibitemShut {NoStop}%
\bibitem [{\citenamefont {Kniehl}\ \emph {et~al.}(2005)\citenamefont {Kniehl},
  \citenamefont {Kramer}, \citenamefont {Schienbein},\ and\ \citenamefont
  {Spiesberger}}]{Kniehl:2004fy}%
  \BibitemOpen
  \bibfield  {author} {\bibinfo {author} {\bibfnamefont {B.~A.}\ \bibnamefont
  {Kniehl}}, \bibinfo {author} {\bibfnamefont {G.}~\bibnamefont {Kramer}},
  \bibinfo {author} {\bibfnamefont {I.}~\bibnamefont {Schienbein}},\ and\
  \bibinfo {author} {\bibfnamefont {H.}~\bibnamefont {Spiesberger}},\
  }\bibfield  {title} {\bibinfo {title} {{Inclusive D*+- production in p anti-p
  collisions with massive charm quarks}},\ }\href
  {https://doi.org/10.1103/PhysRevD.71.014018} {\bibfield  {journal} {\bibinfo
  {journal} {Phys. Rev. D}\ }\textbf {\bibinfo {volume} {71}},\ \bibinfo
  {pages} {014018} (\bibinfo {year} {2005})},\ \Eprint
  {https://arxiv.org/abs/hep-ph/0410289} {arXiv:hep-ph/0410289} \BibitemShut
  {NoStop}%
\bibitem [{\citenamefont {Helenius}\ and\ \citenamefont
  {Paukkunen}(2018)}]{Helenius:2018uul}%
  \BibitemOpen
  \bibfield  {author} {\bibinfo {author} {\bibfnamefont {I.}~\bibnamefont
  {Helenius}}\ and\ \bibinfo {author} {\bibfnamefont {H.}~\bibnamefont
  {Paukkunen}},\ }\bibfield  {title} {\bibinfo {title} {{Revisiting the D-meson
  hadroproduction in general-mass variable flavour number scheme}},\ }\href
  {https://doi.org/10.1007/JHEP05(2018)196} {\bibfield  {journal} {\bibinfo
  {journal} {JHEP}\ }\textbf {\bibinfo {volume} {05}},\ \bibinfo {pages}
  {196}},\ \Eprint {https://arxiv.org/abs/1804.03557} {arXiv:1804.03557
  [hep-ph]} \BibitemShut {NoStop}%
\bibitem [{\citenamefont {Cacciari}\ \emph {et~al.}(2005)\citenamefont
  {Cacciari}, \citenamefont {Nason},\ and\ \citenamefont
  {Vogt}}]{Cacciari:2005rk}%
  \BibitemOpen
  \bibfield  {author} {\bibinfo {author} {\bibfnamefont {M.}~\bibnamefont
  {Cacciari}}, \bibinfo {author} {\bibfnamefont {P.}~\bibnamefont {Nason}},\
  and\ \bibinfo {author} {\bibfnamefont {R.}~\bibnamefont {Vogt}},\ }\bibfield
  {title} {\bibinfo {title} {{QCD predictions for charm and bottom production
  at RHIC}},\ }\href {https://doi.org/10.1103/PhysRevLett.95.122001} {\bibfield
   {journal} {\bibinfo  {journal} {Phys. Rev. Lett.}\ }\textbf {\bibinfo
  {volume} {95}},\ \bibinfo {pages} {122001} (\bibinfo {year} {2005})},\
  \Eprint {https://arxiv.org/abs/hep-ph/0502203} {arXiv:hep-ph/0502203}
  \BibitemShut {NoStop}%
\bibitem [{\citenamefont {Das}\ \emph {et~al.}(2010)\citenamefont {Das},
  \citenamefont {Alam},\ and\ \citenamefont {Mohanty}}]{Das:2010tj}%
  \BibitemOpen
  \bibfield  {author} {\bibinfo {author} {\bibfnamefont {S.~K.}\ \bibnamefont
  {Das}}, \bibinfo {author} {\bibfnamefont {J.-e.}\ \bibnamefont {Alam}},\ and\
  \bibinfo {author} {\bibfnamefont {P.}~\bibnamefont {Mohanty}},\ }\bibfield
  {title} {\bibinfo {title} {{Dragging Heavy Quarks in Quark Gluon Plasma at
  the Large Hadron Collider}},\ }\href
  {https://doi.org/10.1103/PhysRevC.82.014908} {\bibfield  {journal} {\bibinfo
  {journal} {Phys. Rev. C}\ }\textbf {\bibinfo {volume} {82}},\ \bibinfo
  {pages} {014908} (\bibinfo {year} {2010})},\ \Eprint
  {https://arxiv.org/abs/1003.5508} {arXiv:1003.5508 [nucl-th]} \BibitemShut
  {NoStop}%
\bibitem [{\citenamefont {He}\ \emph {et~al.}(2013)\citenamefont {He},
  \citenamefont {Fries},\ and\ \citenamefont {Rapp}}]{He:2012df}%
  \BibitemOpen
  \bibfield  {author} {\bibinfo {author} {\bibfnamefont {M.}~\bibnamefont
  {He}}, \bibinfo {author} {\bibfnamefont {R.~J.}\ \bibnamefont {Fries}},\ and\
  \bibinfo {author} {\bibfnamefont {R.}~\bibnamefont {Rapp}},\ }\bibfield
  {title} {\bibinfo {title} {{$\mathbf{D_s}$-Meson as Quantitative Probe of
  Diffusion and Hadronization in Nuclear Collisions}},\ }\href
  {https://doi.org/10.1103/PhysRevLett.110.112301} {\bibfield  {journal}
  {\bibinfo  {journal} {Phys. Rev. Lett.}\ }\textbf {\bibinfo {volume} {110}},\
  \bibinfo {pages} {112301} (\bibinfo {year} {2013})},\ \Eprint
  {https://arxiv.org/abs/1204.4442} {arXiv:1204.4442 [nucl-th]} \BibitemShut
  {NoStop}%
\bibitem [{\citenamefont {Cao}\ \emph {et~al.}(2015)\citenamefont {Cao},
  \citenamefont {Qin},\ and\ \citenamefont {Bass}}]{Cao:2015hia}%
  \BibitemOpen
  \bibfield  {author} {\bibinfo {author} {\bibfnamefont {S.}~\bibnamefont
  {Cao}}, \bibinfo {author} {\bibfnamefont {G.-Y.}\ \bibnamefont {Qin}},\ and\
  \bibinfo {author} {\bibfnamefont {S.~A.}\ \bibnamefont {Bass}},\ }\bibfield
  {title} {\bibinfo {title} {{Energy loss, hadronization and hadronic
  interactions of heavy flavors in relativistic heavy-ion collisions}},\ }\href
  {https://doi.org/10.1103/PhysRevC.92.024907} {\bibfield  {journal} {\bibinfo
  {journal} {Phys. Rev. C}\ }\textbf {\bibinfo {volume} {92}},\ \bibinfo
  {pages} {024907} (\bibinfo {year} {2015})},\ \Eprint
  {https://arxiv.org/abs/1505.01413} {arXiv:1505.01413 [nucl-th]} \BibitemShut
  {NoStop}%
\bibitem [{\citenamefont {Lang}\ \emph {et~al.}(2016)\citenamefont {Lang},
  \citenamefont {van Hees}, \citenamefont {Steinheimer}, \citenamefont
  {Inghirami},\ and\ \citenamefont {Bleicher}}]{Lang:2012nqy}%
  \BibitemOpen
  \bibfield  {author} {\bibinfo {author} {\bibfnamefont {T.}~\bibnamefont
  {Lang}}, \bibinfo {author} {\bibfnamefont {H.}~\bibnamefont {van Hees}},
  \bibinfo {author} {\bibfnamefont {J.}~\bibnamefont {Steinheimer}}, \bibinfo
  {author} {\bibfnamefont {G.}~\bibnamefont {Inghirami}},\ and\ \bibinfo
  {author} {\bibfnamefont {M.}~\bibnamefont {Bleicher}},\ }\bibfield  {title}
  {\bibinfo {title} {{Heavy quark transport in heavy ion collisions at energies
  available at the BNL Relativistic Heavy Ion Collider and at the CERN Large
  Hadron Collider within the UrQMD hybrid model}},\ }\href
  {https://doi.org/10.1103/PhysRevC.93.014901} {\bibfield  {journal} {\bibinfo
  {journal} {Phys. Rev. C}\ }\textbf {\bibinfo {volume} {93}},\ \bibinfo
  {pages} {014901} (\bibinfo {year} {2016})},\ \Eprint
  {https://arxiv.org/abs/1211.6912} {arXiv:1211.6912 [hep-ph]} \BibitemShut
  {NoStop}%
\bibitem [{\citenamefont {Fochler}\ \emph {et~al.}(2010)\citenamefont
  {Fochler}, \citenamefont {Xu},\ and\ \citenamefont
  {Greiner}}]{Fochler:2010wn}%
  \BibitemOpen
  \bibfield  {author} {\bibinfo {author} {\bibfnamefont {O.}~\bibnamefont
  {Fochler}}, \bibinfo {author} {\bibfnamefont {Z.}~\bibnamefont {Xu}},\ and\
  \bibinfo {author} {\bibfnamefont {C.}~\bibnamefont {Greiner}},\ }\bibfield
  {title} {\bibinfo {title} {{Energy loss in a partonic transport model
  including bremsstrahlung processes}},\ }\href
  {https://doi.org/10.1103/PhysRevC.82.024907} {\bibfield  {journal} {\bibinfo
  {journal} {Phys. Rev. C}\ }\textbf {\bibinfo {volume} {82}},\ \bibinfo
  {pages} {024907} (\bibinfo {year} {2010})},\ \Eprint
  {https://arxiv.org/abs/1003.4380} {arXiv:1003.4380 [hep-ph]} \BibitemShut
  {NoStop}%
\bibitem [{\citenamefont {Djordjevic}\ and\ \citenamefont
  {Djordjevic}(2014)}]{Djordjevic:2013xoa}%
  \BibitemOpen
  \bibfield  {author} {\bibinfo {author} {\bibfnamefont {M.}~\bibnamefont
  {Djordjevic}}\ and\ \bibinfo {author} {\bibfnamefont {M.}~\bibnamefont
  {Djordjevic}},\ }\bibfield  {title} {\bibinfo {title} {{LHC jet suppression
  of light and heavy flavor observables}},\ }\href
  {https://doi.org/10.1016/j.physletb.2014.05.053} {\bibfield  {journal}
  {\bibinfo  {journal} {Phys. Lett. B}\ }\textbf {\bibinfo {volume} {734}},\
  \bibinfo {pages} {286} (\bibinfo {year} {2014})},\ \Eprint
  {https://arxiv.org/abs/1307.4098} {arXiv:1307.4098 [hep-ph]} \BibitemShut
  {NoStop}%
\bibitem [{\citenamefont {Xu}\ \emph {et~al.}(2016)\citenamefont {Xu},
  \citenamefont {Liao},\ and\ \citenamefont {Gyulassy}}]{Xu:2015bbz}%
  \BibitemOpen
  \bibfield  {author} {\bibinfo {author} {\bibfnamefont {J.}~\bibnamefont
  {Xu}}, \bibinfo {author} {\bibfnamefont {J.}~\bibnamefont {Liao}},\ and\
  \bibinfo {author} {\bibfnamefont {M.}~\bibnamefont {Gyulassy}},\ }\bibfield
  {title} {\bibinfo {title} {{Bridging Soft-Hard Transport Properties of
  Quark-Gluon Plasmas with CUJET3.0}},\ }\href
  {https://doi.org/10.1007/JHEP02(2016)169} {\bibfield  {journal} {\bibinfo
  {journal} {JHEP}\ }\textbf {\bibinfo {volume} {02}},\ \bibinfo {pages}
  {169}},\ \Eprint {https://arxiv.org/abs/1508.00552} {arXiv:1508.00552
  [hep-ph]} \BibitemShut {NoStop}%
\bibitem [{\citenamefont {Song}\ \emph {et~al.}(2016)\citenamefont {Song},
  \citenamefont {Berrehrah}, \citenamefont {Cabrera}, \citenamefont {Cassing},\
  and\ \citenamefont {Bratkovskaya}}]{Song:2015ykw}%
  \BibitemOpen
  \bibfield  {author} {\bibinfo {author} {\bibfnamefont {T.}~\bibnamefont
  {Song}}, \bibinfo {author} {\bibfnamefont {H.}~\bibnamefont {Berrehrah}},
  \bibinfo {author} {\bibfnamefont {D.}~\bibnamefont {Cabrera}}, \bibinfo
  {author} {\bibfnamefont {W.}~\bibnamefont {Cassing}},\ and\ \bibinfo {author}
  {\bibfnamefont {E.}~\bibnamefont {Bratkovskaya}},\ }\bibfield  {title}
  {\bibinfo {title} {{Charm production in Pb + Pb collisions at energies
  available at the CERN Large Hadron Collider}},\ }\href
  {https://doi.org/10.1103/PhysRevC.93.034906} {\bibfield  {journal} {\bibinfo
  {journal} {Phys. Rev. C}\ }\textbf {\bibinfo {volume} {93}},\ \bibinfo
  {pages} {034906} (\bibinfo {year} {2016})},\ \Eprint
  {https://arxiv.org/abs/1512.00891} {arXiv:1512.00891 [nucl-th]} \BibitemShut
  {NoStop}%
\bibitem [{\citenamefont {Cao}\ \emph {et~al.}(2016)\citenamefont {Cao},
  \citenamefont {Luo}, \citenamefont {Qin},\ and\ \citenamefont
  {Wang}}]{Cao:2016gvr}%
  \BibitemOpen
  \bibfield  {author} {\bibinfo {author} {\bibfnamefont {S.}~\bibnamefont
  {Cao}}, \bibinfo {author} {\bibfnamefont {T.}~\bibnamefont {Luo}}, \bibinfo
  {author} {\bibfnamefont {G.-Y.}\ \bibnamefont {Qin}},\ and\ \bibinfo {author}
  {\bibfnamefont {X.-N.}\ \bibnamefont {Wang}},\ }\bibfield  {title} {\bibinfo
  {title} {{Linearized Boltzmann transport model for jet propagation in the
  quark-gluon plasma: Heavy quark evolution}},\ }\href
  {https://doi.org/10.1103/PhysRevC.94.014909} {\bibfield  {journal} {\bibinfo
  {journal} {Phys. Rev. C}\ }\textbf {\bibinfo {volume} {94}},\ \bibinfo
  {pages} {014909} (\bibinfo {year} {2016})},\ \Eprint
  {https://arxiv.org/abs/1605.06447} {arXiv:1605.06447 [nucl-th]} \BibitemShut
  {NoStop}%
\bibitem [{\citenamefont {Zhang}\ \emph {et~al.}(2023)\citenamefont {Zhang},
  \citenamefont {Zheng}, \citenamefont {Shi},\ and\ \citenamefont
  {Lin}}]{Zhang:2022fum}%
  \BibitemOpen
  \bibfield  {author} {\bibinfo {author} {\bibfnamefont {C.}~\bibnamefont
  {Zhang}}, \bibinfo {author} {\bibfnamefont {L.}~\bibnamefont {Zheng}},
  \bibinfo {author} {\bibfnamefont {S.}~\bibnamefont {Shi}},\ and\ \bibinfo
  {author} {\bibfnamefont {Z.-W.}\ \bibnamefont {Lin}},\ }\bibfield  {title}
  {\bibinfo {title} {{Resolving the RpA and v2 puzzle of D0 mesons in
  p{\ensuremath{-}}Pb collisions at the LHC}},\ }\href
  {https://doi.org/10.1016/j.physletb.2023.138219} {\bibfield  {journal}
  {\bibinfo  {journal} {Phys. Lett. B}\ }\textbf {\bibinfo {volume} {846}},\
  \bibinfo {pages} {138219} (\bibinfo {year} {2023})},\ \Eprint
  {https://arxiv.org/abs/2210.07767} {arXiv:2210.07767 [nucl-th]} \BibitemShut
  {NoStop}%
\bibitem [{\citenamefont {Zhang}\ \emph {et~al.}(2024)\citenamefont {Zhang},
  \citenamefont {Zheng}, \citenamefont {Shi},\ and\ \citenamefont
  {Lin}}]{Zhang:2024zga}%
  \BibitemOpen
  \bibfield  {author} {\bibinfo {author} {\bibfnamefont {C.}~\bibnamefont
  {Zhang}}, \bibinfo {author} {\bibfnamefont {L.}~\bibnamefont {Zheng}},
  \bibinfo {author} {\bibfnamefont {S.}~\bibnamefont {Shi}},\ and\ \bibinfo
  {author} {\bibfnamefont {Z.-W.}\ \bibnamefont {Lin}},\ }\bibfield  {title}
  {\bibinfo {title} {{Investigating $D^0$ meson production in p-Pb collisions
  at 5.02 TeV with a multi-phase transport model}},\ }\href
  {https://doi.org/10.1140/epjc/s10052-024-13265-9} {\bibfield  {journal}
  {\bibinfo  {journal} {Eur. Phys. J. C}\ }\textbf {\bibinfo {volume} {84}},\
  \bibinfo {pages} {942} (\bibinfo {year} {2024})},\ \Eprint
  {https://arxiv.org/abs/2403.06099} {arXiv:2403.06099 [nucl-th]} \BibitemShut
  {NoStop}%
\bibitem [{\citenamefont {Lin}\ \emph {et~al.}(2005)\citenamefont {Lin},
  \citenamefont {Ko}, \citenamefont {Li}, \citenamefont {Zhang},\ and\
  \citenamefont {Pal}}]{Lin:2004en}%
  \BibitemOpen
  \bibfield  {author} {\bibinfo {author} {\bibfnamefont {Z.-W.}\ \bibnamefont
  {Lin}}, \bibinfo {author} {\bibfnamefont {C.~M.}\ \bibnamefont {Ko}},
  \bibinfo {author} {\bibfnamefont {B.-A.}\ \bibnamefont {Li}}, \bibinfo
  {author} {\bibfnamefont {B.}~\bibnamefont {Zhang}},\ and\ \bibinfo {author}
  {\bibfnamefont {S.}~\bibnamefont {Pal}},\ }\bibfield  {title} {\bibinfo
  {title} {{A Multi-phase transport model for relativistic heavy ion
  collisions}},\ }\href {https://doi.org/10.1103/PhysRevC.72.064901} {\bibfield
   {journal} {\bibinfo  {journal} {Phys. Rev. C}\ }\textbf {\bibinfo {volume}
  {72}},\ \bibinfo {pages} {064901} (\bibinfo {year} {2005})},\ \Eprint
  {https://arxiv.org/abs/nucl-th/0411110} {arXiv:nucl-th/0411110} \BibitemShut
  {NoStop}%
\bibitem [{\citenamefont {He}\ and\ \citenamefont {Lin}(2017)}]{He:2017tla}%
  \BibitemOpen
  \bibfield  {author} {\bibinfo {author} {\bibfnamefont {Y.}~\bibnamefont
  {He}}\ and\ \bibinfo {author} {\bibfnamefont {Z.-W.}\ \bibnamefont {Lin}},\
  }\bibfield  {title} {\bibinfo {title} {{Improved Quark Coalescence for a
  Multi-Phase Transport Model}},\ }\href
  {https://doi.org/10.1103/PhysRevC.96.014910} {\bibfield  {journal} {\bibinfo
  {journal} {Phys. Rev. C}\ }\textbf {\bibinfo {volume} {96}},\ \bibinfo
  {pages} {014910} (\bibinfo {year} {2017})},\ \Eprint
  {https://arxiv.org/abs/1703.02673} {arXiv:1703.02673 [nucl-th]} \BibitemShut
  {NoStop}%
\bibitem [{\citenamefont {Zheng}\ \emph {et~al.}(2020)\citenamefont {Zheng},
  \citenamefont {Zhang}, \citenamefont {Shi},\ and\ \citenamefont
  {Lin}}]{Zheng:2019alz}%
  \BibitemOpen
  \bibfield  {author} {\bibinfo {author} {\bibfnamefont {L.}~\bibnamefont
  {Zheng}}, \bibinfo {author} {\bibfnamefont {C.}~\bibnamefont {Zhang}},
  \bibinfo {author} {\bibfnamefont {S.~S.}\ \bibnamefont {Shi}},\ and\ \bibinfo
  {author} {\bibfnamefont {Z.~W.}\ \bibnamefont {Lin}},\ }\bibfield  {title}
  {\bibinfo {title} {{Improvement of heavy flavor production in a multiphase
  transport model updated with modern nuclear parton distribution functions}},\
  }\href {https://doi.org/10.1103/PhysRevC.101.034905} {\bibfield  {journal}
  {\bibinfo  {journal} {Phys. Rev. C}\ }\textbf {\bibinfo {volume} {101}},\
  \bibinfo {pages} {034905} (\bibinfo {year} {2020})},\ \Eprint
  {https://arxiv.org/abs/1909.07191} {arXiv:1909.07191 [nucl-th]} \BibitemShut
  {NoStop}%
\bibitem [{\citenamefont {Zhang}\ \emph {et~al.}(2019)\citenamefont {Zhang},
  \citenamefont {Zheng}, \citenamefont {Liu}, \citenamefont {Shi},\ and\
  \citenamefont {Lin}}]{Zhang:2019utb}%
  \BibitemOpen
  \bibfield  {author} {\bibinfo {author} {\bibfnamefont {C.}~\bibnamefont
  {Zhang}}, \bibinfo {author} {\bibfnamefont {L.}~\bibnamefont {Zheng}},
  \bibinfo {author} {\bibfnamefont {F.}~\bibnamefont {Liu}}, \bibinfo {author}
  {\bibfnamefont {S.}~\bibnamefont {Shi}},\ and\ \bibinfo {author}
  {\bibfnamefont {Z.-W.}\ \bibnamefont {Lin}},\ }\bibfield  {title} {\bibinfo
  {title} {{Update of a multiphase transport model with modern parton
  distribution functions and nuclear shadowing}},\ }\href
  {https://doi.org/10.1103/PhysRevC.99.064906} {\bibfield  {journal} {\bibinfo
  {journal} {Phys. Rev. C}\ }\textbf {\bibinfo {volume} {99}},\ \bibinfo
  {pages} {064906} (\bibinfo {year} {2019})},\ \Eprint
  {https://arxiv.org/abs/1903.03292} {arXiv:1903.03292 [nucl-th]} \BibitemShut
  {NoStop}%
\bibitem [{\citenamefont {Lin}\ and\ \citenamefont
  {Zheng}(2021)}]{Lin:2021mdn}%
  \BibitemOpen
  \bibfield  {author} {\bibinfo {author} {\bibfnamefont {Z.-W.}\ \bibnamefont
  {Lin}}\ and\ \bibinfo {author} {\bibfnamefont {L.}~\bibnamefont {Zheng}},\
  }\bibfield  {title} {\bibinfo {title} {{Further developments of a multi-phase
  transport model for relativistic nuclear collisions}},\ }\href
  {https://doi.org/10.1007/s41365-021-00944-5} {\bibfield  {journal} {\bibinfo
  {journal} {Nucl. Sci. Tech.}\ }\textbf {\bibinfo {volume} {32}},\ \bibinfo
  {pages} {113} (\bibinfo {year} {2021})},\ \Eprint
  {https://arxiv.org/abs/2110.02989} {arXiv:2110.02989 [nucl-th]} \BibitemShut
  {NoStop}%
\bibitem [{\citenamefont {Wang}\ and\ \citenamefont
  {Gyulassy}(1991)}]{Wang:1991hta}%
  \BibitemOpen
  \bibfield  {author} {\bibinfo {author} {\bibfnamefont {X.-N.}\ \bibnamefont
  {Wang}}\ and\ \bibinfo {author} {\bibfnamefont {M.}~\bibnamefont
  {Gyulassy}},\ }\bibfield  {title} {\bibinfo {title} {{HIJING: A Monte Carlo
  model for multiple jet production in p p, p A and A A collisions}},\ }\href
  {https://doi.org/10.1103/PhysRevD.44.3501} {\bibfield  {journal} {\bibinfo
  {journal} {Phys. Rev. D}\ }\textbf {\bibinfo {volume} {44}},\ \bibinfo
  {pages} {3501} (\bibinfo {year} {1991})}\BibitemShut {NoStop}%
\bibitem [{\citenamefont {Zhang}(1998)}]{Zhang:1997ej}%
  \BibitemOpen
  \bibfield  {author} {\bibinfo {author} {\bibfnamefont {B.}~\bibnamefont
  {Zhang}},\ }\bibfield  {title} {\bibinfo {title} {{ZPC 1.0.1: A Parton
  cascade for ultrarelativistic heavy ion collisions}},\ }\href
  {https://doi.org/10.1016/S0010-4655(98)00010-1} {\bibfield  {journal}
  {\bibinfo  {journal} {Comput. Phys. Commun.}\ }\textbf {\bibinfo {volume}
  {109}},\ \bibinfo {pages} {193} (\bibinfo {year} {1998})},\ \Eprint
  {https://arxiv.org/abs/nucl-th/9709009} {arXiv:nucl-th/9709009} \BibitemShut
  {NoStop}%
\bibitem [{\citenamefont {Cronin}\ \emph {et~al.}(1975)\citenamefont {Cronin},
  \citenamefont {Frisch}, \citenamefont {Shochet}, \citenamefont {Boymond},
  \citenamefont {Mermod}, \citenamefont {Piroue},\ and\ \citenamefont
  {Sumner}}]{Cronin:1974zm}%
  \BibitemOpen
  \bibfield  {author} {\bibinfo {author} {\bibfnamefont {J.~W.}\ \bibnamefont
  {Cronin}}, \bibinfo {author} {\bibfnamefont {H.~J.}\ \bibnamefont {Frisch}},
  \bibinfo {author} {\bibfnamefont {M.~J.}\ \bibnamefont {Shochet}}, \bibinfo
  {author} {\bibfnamefont {J.~P.}\ \bibnamefont {Boymond}}, \bibinfo {author}
  {\bibfnamefont {R.}~\bibnamefont {Mermod}}, \bibinfo {author} {\bibfnamefont
  {P.~A.}\ \bibnamefont {Piroue}},\ and\ \bibinfo {author} {\bibfnamefont
  {R.~L.}\ \bibnamefont {Sumner}} (\bibinfo {collaboration} {E100}),\
  }\bibfield  {title} {\bibinfo {title} {{Production of hadrons with large
  transverse momentum at 200, 300, and 400 GeV}},\ }\href
  {https://doi.org/10.1103/PhysRevD.11.3105} {\bibfield  {journal} {\bibinfo
  {journal} {Phys. Rev. D}\ }\textbf {\bibinfo {volume} {11}},\ \bibinfo
  {pages} {3105} (\bibinfo {year} {1975})}\BibitemShut {NoStop}%
\bibitem [{\citenamefont {Accardi}(2002)}]{Accardi:2002ik}%
  \BibitemOpen
  \bibfield  {author} {\bibinfo {author} {\bibfnamefont {A.}~\bibnamefont
  {Accardi}},\ }\bibfield  {title} {\bibinfo {title} {{Cronin effect in proton
  nucleus collisions: A Survey of theoretical models}},\ }\href@noop {} {\
  (\bibinfo {year} {2002})},\ \Eprint {https://arxiv.org/abs/hep-ph/0212148}
  {arXiv:hep-ph/0212148} \BibitemShut {NoStop}%
\bibitem [{\citenamefont {Vitev}\ \emph {et~al.}(2006)\citenamefont {Vitev},
  \citenamefont {Goldman}, \citenamefont {Johnson},\ and\ \citenamefont
  {Qiu}}]{Vitev:2006bi}%
  \BibitemOpen
  \bibfield  {author} {\bibinfo {author} {\bibfnamefont {I.}~\bibnamefont
  {Vitev}}, \bibinfo {author} {\bibfnamefont {J.~T.}\ \bibnamefont {Goldman}},
  \bibinfo {author} {\bibfnamefont {M.~B.}\ \bibnamefont {Johnson}},\ and\
  \bibinfo {author} {\bibfnamefont {J.~W.}\ \bibnamefont {Qiu}},\ }\bibfield
  {title} {\bibinfo {title} {{Open charm tomography of cold nuclear matter}},\
  }\href {https://doi.org/10.1103/PhysRevD.74.054010} {\bibfield  {journal}
  {\bibinfo  {journal} {Phys. Rev. D}\ }\textbf {\bibinfo {volume} {74}},\
  \bibinfo {pages} {054010} (\bibinfo {year} {2006})},\ \Eprint
  {https://arxiv.org/abs/hep-ph/0605200} {arXiv:hep-ph/0605200} \BibitemShut
  {NoStop}%
\bibitem [{\citenamefont {Andersson}\ \emph
  {et~al.}(1983{\natexlab{a}})\citenamefont {Andersson}, \citenamefont
  {Gustafson}, \citenamefont {Ingelman},\ and\ \citenamefont
  {Sjostrand}}]{Andersson:1983ia}%
  \BibitemOpen
  \bibfield  {author} {\bibinfo {author} {\bibfnamefont {B.}~\bibnamefont
  {Andersson}}, \bibinfo {author} {\bibfnamefont {G.}~\bibnamefont
  {Gustafson}}, \bibinfo {author} {\bibfnamefont {G.}~\bibnamefont
  {Ingelman}},\ and\ \bibinfo {author} {\bibfnamefont {T.}~\bibnamefont
  {Sjostrand}},\ }\bibfield  {title} {\bibinfo {title} {{Parton Fragmentation
  and String Dynamics}},\ }\href {https://doi.org/10.1016/0370-1573(83)90080-7}
  {\bibfield  {journal} {\bibinfo  {journal} {Phys. Rept.}\ }\textbf {\bibinfo
  {volume} {97}},\ \bibinfo {pages} {31} (\bibinfo {year}
  {1983}{\natexlab{a}})}\BibitemShut {NoStop}%
\bibitem [{\citenamefont {Andersson}\ \emph
  {et~al.}(1983{\natexlab{b}})\citenamefont {Andersson}, \citenamefont
  {Gustafson},\ and\ \citenamefont {Soderberg}}]{Andersson:1983jt}%
  \BibitemOpen
  \bibfield  {author} {\bibinfo {author} {\bibfnamefont {B.}~\bibnamefont
  {Andersson}}, \bibinfo {author} {\bibfnamefont {G.}~\bibnamefont
  {Gustafson}},\ and\ \bibinfo {author} {\bibfnamefont {B.}~\bibnamefont
  {Soderberg}},\ }\bibfield  {title} {\bibinfo {title} {{A General Model for
  Jet Fragmentation}},\ }\href {https://doi.org/10.1007/BF01407824} {\bibfield
  {journal} {\bibinfo  {journal} {Z. Phys. C}\ }\textbf {\bibinfo {volume}
  {20}},\ \bibinfo {pages} {317} (\bibinfo {year}
  {1983}{\natexlab{b}})}\BibitemShut {NoStop}%
\bibitem [{\citenamefont {Cacciari}\ and\ \citenamefont
  {Gardi}(2003)}]{Cacciari:2002xb}%
  \BibitemOpen
  \bibfield  {author} {\bibinfo {author} {\bibfnamefont {M.}~\bibnamefont
  {Cacciari}}\ and\ \bibinfo {author} {\bibfnamefont {E.}~\bibnamefont
  {Gardi}},\ }\bibfield  {title} {\bibinfo {title} {{Heavy quark
  fragmentation}},\ }\href {https://doi.org/10.1016/S0550-3213(03)00435-8}
  {\bibfield  {journal} {\bibinfo  {journal} {Nucl. Phys. B}\ }\textbf
  {\bibinfo {volume} {664}},\ \bibinfo {pages} {299} (\bibinfo {year}
  {2003})},\ \Eprint {https://arxiv.org/abs/hep-ph/0301047}
  {arXiv:hep-ph/0301047} \BibitemShut {NoStop}%
\bibitem [{\citenamefont {Moosavi~Nejad}\ and\ \citenamefont
  {Armat}(2013)}]{MoosaviNejad:2013ssd}%
  \BibitemOpen
  \bibfield  {author} {\bibinfo {author} {\bibfnamefont {S.~M.}\ \bibnamefont
  {Moosavi~Nejad}}\ and\ \bibinfo {author} {\bibfnamefont {A.}~\bibnamefont
  {Armat}},\ }\bibfield  {title} {\bibinfo {title} {{Heavy quark perturbative
  QCD fragmentation functions in the presence of hadron mass}},\ }\href
  {https://doi.org/10.1140/epjp/i2013-13121-2} {\bibfield  {journal} {\bibinfo
  {journal} {Eur. Phys. J. Plus}\ }\textbf {\bibinfo {volume} {128}},\ \bibinfo
  {pages} {121} (\bibinfo {year} {2013})},\ \Eprint
  {https://arxiv.org/abs/1307.6351} {arXiv:1307.6351 [hep-ph]} \BibitemShut
  {NoStop}%
\bibitem [{\citenamefont {Mele}\ and\ \citenamefont
  {Nason}(1991)}]{Mele:1990cw}%
  \BibitemOpen
  \bibfield  {author} {\bibinfo {author} {\bibfnamefont {B.}~\bibnamefont
  {Mele}}\ and\ \bibinfo {author} {\bibfnamefont {P.}~\bibnamefont {Nason}},\
  }\bibfield  {title} {\bibinfo {title} {{The Fragmentation function for heavy
  quarks in QCD}},\ }\href {https://doi.org/10.1016/0550-3213(91)90597-Q}
  {\bibfield  {journal} {\bibinfo  {journal} {Nucl. Phys. B}\ }\textbf
  {\bibinfo {volume} {361}},\ \bibinfo {pages} {626} (\bibinfo {year}
  {1991})},\ \bibinfo {note} {[Erratum: Nucl.Phys.B 921, 841--842
  (2017)]}\BibitemShut {NoStop}%
\bibitem [{\citenamefont {Braaten}\ \emph {et~al.}(1995)\citenamefont
  {Braaten}, \citenamefont {Cheung}, \citenamefont {Fleming},\ and\
  \citenamefont {Yuan}}]{Braaten:1994bz}%
  \BibitemOpen
  \bibfield  {author} {\bibinfo {author} {\bibfnamefont {E.}~\bibnamefont
  {Braaten}}, \bibinfo {author} {\bibfnamefont {K.-m.}\ \bibnamefont {Cheung}},
  \bibinfo {author} {\bibfnamefont {S.}~\bibnamefont {Fleming}},\ and\ \bibinfo
  {author} {\bibfnamefont {T.~C.}\ \bibnamefont {Yuan}},\ }\bibfield  {title}
  {\bibinfo {title} {{Perturbative QCD fragmentation functions as a model for
  heavy quark fragmentation}},\ }\href
  {https://doi.org/10.1103/PhysRevD.51.4819} {\bibfield  {journal} {\bibinfo
  {journal} {Phys. Rev. D}\ }\textbf {\bibinfo {volume} {51}},\ \bibinfo
  {pages} {4819} (\bibinfo {year} {1995})},\ \Eprint
  {https://arxiv.org/abs/hep-ph/9409316} {arXiv:hep-ph/9409316} \BibitemShut
  {NoStop}%
\bibitem [{\citenamefont {Sjostrand}(1994)}]{Sjostrand:1993yb}%
  \BibitemOpen
  \bibfield  {author} {\bibinfo {author} {\bibfnamefont {T.}~\bibnamefont
  {Sjostrand}},\ }\bibfield  {title} {\bibinfo {title} {{High-energy physics
  event generation with PYTHIA 5.7 and JETSET 7.4}},\ }\href
  {https://doi.org/10.1016/0010-4655(94)90132-5} {\bibfield  {journal}
  {\bibinfo  {journal} {Comput. Phys. Commun.}\ }\textbf {\bibinfo {volume}
  {82}},\ \bibinfo {pages} {74} (\bibinfo {year} {1994})}\BibitemShut {NoStop}%
\bibitem [{\citenamefont {Zhao}\ \emph
  {et~al.}(2024{\natexlab{a}})\citenamefont {Zhao}, \citenamefont {Zhou},
  \citenamefont {Lin}, \citenamefont {Zhang},\ and\ \citenamefont
  {Ma}}]{Zhao:2024feh}%
  \BibitemOpen
  \bibfield  {author} {\bibinfo {author} {\bibfnamefont {X.-L.}\ \bibnamefont
  {Zhao}}, \bibinfo {author} {\bibfnamefont {Y.}~\bibnamefont {Zhou}}, \bibinfo
  {author} {\bibfnamefont {Z.-W.}\ \bibnamefont {Lin}}, \bibinfo {author}
  {\bibfnamefont {C.}~\bibnamefont {Zhang}},\ and\ \bibinfo {author}
  {\bibfnamefont {G.-L.}\ \bibnamefont {Ma}},\ }\bibfield  {title} {\bibinfo
  {title} {{Nuclear cluster structure effect in $^{16}$O+$^{16}$O collisions at
  the top RHIC energy}},\ }\href@noop {} {\  (\bibinfo {year}
  {2024}{\natexlab{a}})},\ \Eprint {https://arxiv.org/abs/2404.09780}
  {arXiv:2404.09780 [nucl-th]} \BibitemShut {NoStop}%
\bibitem [{\citenamefont {Zhao}\ \emph
  {et~al.}(2024{\natexlab{b}})\citenamefont {Zhao} \emph
  {et~al.}}]{Zhao:2023nrz}%
  \BibitemOpen
  \bibfield  {author} {\bibinfo {author} {\bibfnamefont {J.}~\bibnamefont
  {Zhao}} \emph {et~al.},\ }\bibfield  {title} {\bibinfo {title}
  {{Hadronization of heavy quarks}},\ }\href
  {https://doi.org/10.1103/PhysRevC.109.054912} {\bibfield  {journal} {\bibinfo
   {journal} {Phys. Rev. C}\ }\textbf {\bibinfo {volume} {109}},\ \bibinfo
  {pages} {054912} (\bibinfo {year} {2024}{\natexlab{b}})},\ \Eprint
  {https://arxiv.org/abs/2311.10621} {arXiv:2311.10621 [hep-ph]} \BibitemShut
  {NoStop}%
\bibitem [{\citenamefont {Hwa}\ and\ \citenamefont {Yang}(2004)}]{Hwa:2004ng}%
  \BibitemOpen
  \bibfield  {author} {\bibinfo {author} {\bibfnamefont {R.~C.}\ \bibnamefont
  {Hwa}}\ and\ \bibinfo {author} {\bibfnamefont {C.~B.}\ \bibnamefont {Yang}},\
  }\bibfield  {title} {\bibinfo {title} {{Recombination of shower partons at
  high p(T) in heavy ion collisions}},\ }\href
  {https://doi.org/10.1103/PhysRevC.70.024905} {\bibfield  {journal} {\bibinfo
  {journal} {Phys. Rev. C}\ }\textbf {\bibinfo {volume} {70}},\ \bibinfo
  {pages} {024905} (\bibinfo {year} {2004})},\ \Eprint
  {https://arxiv.org/abs/nucl-th/0401001} {arXiv:nucl-th/0401001} \BibitemShut
  {NoStop}%
\bibitem [{\citenamefont {Song}\ and\ \citenamefont
  {Coci}(2022)}]{Song:2021mvc}%
  \BibitemOpen
  \bibfield  {author} {\bibinfo {author} {\bibfnamefont {T.}~\bibnamefont
  {Song}}\ and\ \bibinfo {author} {\bibfnamefont {G.}~\bibnamefont {Coci}},\
  }\bibfield  {title} {\bibinfo {title} {{Prerequisites for heavy quark
  coalescence in heavy-ion collisions}},\ }\href
  {https://doi.org/10.1016/j.nuclphysa.2022.122539} {\bibfield  {journal}
  {\bibinfo  {journal} {Nucl. Phys. A}\ }\textbf {\bibinfo {volume} {1028}},\
  \bibinfo {pages} {122539} (\bibinfo {year} {2022})},\ \Eprint
  {https://arxiv.org/abs/2104.10987} {arXiv:2104.10987 [nucl-th]} \BibitemShut
  {NoStop}%
\bibitem [{\citenamefont {Zhang}\ \emph {et~al.}(2025)\citenamefont {Zhang},
  \citenamefont {Peng}, \citenamefont {Peng},\ and\ \citenamefont
  {Zheng}}]{Zhang:2025pqu}%
  \BibitemOpen
  \bibfield  {author} {\bibinfo {author} {\bibfnamefont {A.-G.}\ \bibnamefont
  {Zhang}}, \bibinfo {author} {\bibfnamefont {X.-Y.}\ \bibnamefont {Peng}},
  \bibinfo {author} {\bibfnamefont {X.}~\bibnamefont {Peng}},\ and\ \bibinfo
  {author} {\bibfnamefont {L.}~\bibnamefont {Zheng}},\ }\bibfield  {title}
  {\bibinfo {title} {{Exploring the multiplicity dependence of the flavor
  hierarchy for hadron production in high-energy pp collisions}},\ }\href
  {https://doi.org/10.1007/s41365-025-01728-x} {\bibfield  {journal} {\bibinfo
  {journal} {Nucl. Sci. Tech.}\ }\textbf {\bibinfo {volume} {36}},\ \bibinfo
  {pages} {134} (\bibinfo {year} {2025})},\ \Eprint
  {https://arxiv.org/abs/2503.10157} {arXiv:2503.10157 [hep-ph]} \BibitemShut
  {NoStop}%
\end{thebibliography}%

\end{document}